\documentclass[preprint]{aastex63} 

\newcommand{\code}[1]{\texttt{#1}}
\newcommand{\pathc}[1]{\code{\path{#1}}}
\newcommand{\newalpha}{\alpha_{0.3}}
\newcommand{\newa}{a_{0.1}}
\newcommand{\newmdotdnaught}{\dot{m}_{d,0.1}}
\newcommand{\kappacorr}{\Tilde{\kappa}}
\newcommand{\pcorr}{\Tilde{p}}
\newcommand{\evaprate}{\dot{\Sigma}_z}

\graphicspath{{./}{figures/}}

\usepackage[]{verbatim}

\usepackage{amsmath}
\usepackage{pgf}

\usepackage{enumitem}

\usepackage{physics}

\usepackage{xspace}

\usepackage{CJK}

\newcommand{\brem}{bremsstrahlung\xspace}

\received{January 26, 2022}

\submitjournal{ApJ}

\shorttitle{Analytical Model of Disk Evaporation}
\shortauthors{Cho \& Narayan}

\begin{document}
\begin{CJK}{UTF8}{mj}

\defcitealias{Meyer1994}{MMH94}
\newcommand{\mmh}{\citetalias{Meyer1994}\xspace}
\defcitealias{Meyer2000}{MLMH00}
\newcommand{\mlmh}{\citetalias{Meyer2000}\xspace}

\title{Analytical Model of Disk Evaporation and State Transitions in Accreting Black Holes}

\correspondingauthor{Hyerin Cho}
\email{hyerin.cho@cfa.harvard.edu}
\author[0000-0002-2858-9481]{Hyerin Cho (조혜린)}
\affiliation{Center for Astrophysics $\vert$ Harvard \& Smithsonian, 60 Garden Street, Cambridge, MA 02138, USA}
\affiliation{Black Hole Initiative at Harvard University, 20 Garden Street, Cambridge, MA 02138, USA}

\author{Ramesh Narayan}
\affiliation{Center for Astrophysics $\vert$ Harvard \& Smithsonian, 60 Garden Street, Cambridge, MA 02138, USA}
\affiliation{Black Hole Initiative at Harvard University, 20 Garden Street, Cambridge, MA 02138, USA}

\begin{abstract} 

State transitions in black hole X-ray binaries are likely caused by gas evaporation from a thin accretion disk into a hot corona. 
We present a height-integrated version of this process which is suitable for analytical and numerical studies. With radius $r$ scaled to Schwarzschild units and coronal mass accretion rate $\dot{m}_c$ to Eddington units, the results of the model are independent of black hole mass. State transitions should thus be similar in X-ray binaries and AGN.
The corona solution consists of two power-law segments separated at a break radius $r_b \sim10^3 \,(\alpha/0.3)^{-2}$, where $\alpha$ is the viscosity parameter. Gas evaporates from the disk to the corona for $r>r_b$, and condenses back for $r<r_b$. At $r_b$, $\dot{m}_c$ reaches its maximum, $\dot{m}_{c,{\rm max}} \approx 0.02\, (\alpha/0.3)^3$. If at $r\gg r_b$ the thin disk accretes with $\dot{m}_d < \dot{m}_{c,{\rm max}} $, then the disk evaporates fully before reaching $r_b$, giving the hard state. Otherwise, the disk survives at all radii, giving the thermal state.
While the basic model considers only \brem cooling and viscous heating, we also discuss a more realistic model which includes Compton cooling and direct coronal heating by energy transport from the disk. Solutions are again independent of black hole mass, and $r_b$ remains unchanged. This model predicts strong coronal winds for $r>r_b$, and a $T\sim 5\times 10^8\,{\rm K}$ Compton-cooled corona for $r < r_b$. Two-temperature effects are ignored, but may be important at small radii.
\end{abstract}


\section{Introduction} \label{sec:intro}

Black hole (BH) X-ray binaries have several distinct accretion states (see \citealt{Tanaka1996,McClintock_Remillard_2006,Remillard2006,Done+2007} for reviews). The most notable of these are the thermal (or high/soft) state and the hard (or low/hard) state. The thermal state is characterized by a dominant multitemperature blackbody spectrum in X-rays, consistent with a geometrically thin, optically thick, cool accretion disk, similar to the model described in \citet{Shakura1973,Novikov_Thorne_1973,Frank2002}. The hard state, on the other hand, is characterized by a power-law X-ray spectrum extending beyond 100\,keV, which is attributed to radiation from a geometrically thick, optically thin, hot accretion flow (also known as advection-dominated accretion flow, ADAF, or radiatively inefficient accretion flow, RIAF; \citealt{Narayan_Yi_1994,Narayan1995,Abramowicz+_1995}; see \citealt{Yuan2014} for a review). 

Several BH X-ray binaries (BHBs) have been observed during transitions from the thermal state to the hard state, and vice versa. From these observations, it is clear that the thermal state generally corresponds to a larger luminosity and mass accretion rate $\gtrsim$ a few percent of Eddington, while the hard state corresponds to a lower luminosity and accretion rate \citep{Tanaka1996,Done+2007}, though there is a notable hysteresis in the transitions \citep{Remillard2006,Done+2007}. \citet{Esin1997} explained the thermal and hard spectral states as arising from different accretion disk configurations. According to their model, the thermal state corresponds to a configuration in which a thin accretion disk extends all the way down to the innermost stable circular orbit (ISCO), whereas in the hard state, the thin disk is truncated at some truncation radius outside the ISCO and the flow switches to an ADAF inside that radius. The \citet{Esin1997} model is broadly consistent with observations, though it does not readily explain the hysteresis phenomenon. Although some progress has been made on a theoretical explanation of hysteresis \citep{Meyer-Hofmeister2005,Liu2005}, this topic still remains an open question. It is not the focus of the present paper.

Accreting supermassive BHs (SMBHs) in active galactic nuclei (AGN) again show different spectral states. The spectra of luminous AGN (quasars) are dominated by a big blue bump, consistent with a standard thermal thin accretion disk. Low luminosity AGN (LLAGN), on the other hand, have weak or absent blue bumps \citep{Ho1999}, and appear to be consistent with a model in which the thin disk is restricted to large radii, with accretion at smaller radii occurring via an ADAF (see \citealt{Yuan2014}, and references therein). Compared to BHBs, whose $M\sim10M_\odot$ black holes have short time scales, SMBHs with $M\sim10^6-10^{10}M_\odot$ have much longer time scales. Therefore we are less likely to witness a complete state transition within human time scales. Nevertheless, the class of ``changing look" AGN \citep{LaMassa+2015} might well correspond to objects that are undergoing a thermal-to-hard (or vice versa) state transition \citep[e.g.,][]{Noda_Done_2018}. Interestingly, the state transition (if this is what one is seeing) in changing look AGN seems to occur at roughly the same Eddington-scaled luminosity as the corresponding transition in BHBs. Furthermore, in the hard state in both BHBs and AGN, the truncation radius between the outer thin disk and the inner hot ADAF, when scaled to Schwarzschild units, appears to show a similar dependence on the Eddington-scaled luminosity \citep{Yuan2004}. All this suggests that the state transition phenomenon is a robust feature of BH accretion, and that the basic physics is independent of BH mass.

While the model geometry described in  \citet{Esin1997} is successful in explaining the spectra of the thermal and hard states \citep[see, e.g.,][]{Poutanen2018}, the question of how and why state transitions occur is not yet fully understood. It appears that a necessary ingredient is the presence of a hot magnetized corona above the cold thin disk. In seminal work, \citet{Galeev1979} suggested that amplified magnetic field in the disk causes loop-like magnetic structures to emerge due to buoyancy. In analogy with the solar corona, this will result in a very hot magnetically confined corona above the thin disk. Field amplification in the thin disk was later shown to be from the magneto-rotational instability \citep{Balbus_Hawley_1991}.
Coronae have been widely invoked to explain the observation of intense X-rays from cold accretion disks \citep[e.g.,][]{Haardt1991,Svensson1994}. Coronae also play an important role in studies of X-ray reflection spectra \citep{Reynolds2014}.
Coronae form spontaneously in MHD simulations of accretion disks, where they are generated by either the Parker instability \citep{Machida2000} or dissipation of magnetic turbulence \citep[e.g.,][]{Jiang2014}. Coronae are also seen in general relativistic MHD simulations of black hole disks \citep[e.g.,][]{DeVilliers2003,Yuan2014}.

Among several models for accretion state transitions, the best-studied is the evaporation model, which was first proposed by \citet[hereafter \citetalias{Meyer1994}]{Meyer1994} for white dwarf accretion, and later adapted by \citet[hereafter \citetalias{Meyer2000}]{Meyer2000} to explain state transitions in BHBs.
In this model (hereafter ``MM model"), mass and energy are exchanged in the vertical $z$-direction between the thin disk and the overlying corona. The hot corona thermally conducts energy downward to the thin disk and as a result matter in the disk ``evaporates'' and flows into the corona above.  If the evaporation rate at some radius is large enough to remove all the gas in the thin disk, which can happen at sufficiently low mass accretion rates, the disk vanishes entirely and accretion at smaller radii occurs entirely via the corona. \citet{Liu1999} showed that the corona equations are equivalent to the ADAF model in the limit when there is no energy conduction to a cool disk. Therefore, once the thin disk is fully evaporated, the coronal flow automatically becomes a standard ADAF. The MM evaporation model and its subsequent extensions \citep[e.g.,][]{Qiao2009} have been successful in explaining X-ray observations of the state transition in BHBs. However, because of the complex vertical structure of the disk-corona system, as against the simpler height-integrated single layer structure of both the thin disk and ADAF models, studies of the evaporation model have been limited to numerically solving a set of complex differential equations in the vertical direction. 

The goal of the present paper is two-fold. First, using simplifying approximations, we present a height-integrated version of the MM model which is suitable for analytical work as well as for simple numerical calculations. Using this height-integrated model, we analytically reproduce different regimes of the MM evaporation model and explain the relevant physics operating in each regime. We also study how the results depend on parameters such as the viscosity parameter $\alpha$ and the thermal conduction coefficient $\kappa$. Second, we show how to extend the height-integrated model to include additional physical effects beyond the original MM model. As an example, we consider the effect of direct magnetic heating of the corona by turbulent energy transport from the thin disk, and we present analytical results corresponding to this extension of the model. We also outline other extensions that could be explored in the future.

We provide an overview of the basic set-up of our model in Section~\ref{sec:model_setup}, and we present analytical/numerical solutions in Section~\ref{sec:radial_analytic}. We compare the analytical solutions with the original numerical approach of \mmh and \mlmh in Section~\ref{sec:compare_to_meyer}, and describe the implications of our model for state transitions in Section~\ref{sec:truncation_radius}. We include the effects of direct coronal heating and Compton cooling in Section~\ref{sec:mag} and present analytical and numerical results for this version of the model. We discuss applications to observations, comparisons to other models, and the limitations of our approach, in Section~\ref{sec:discussion}, and we summarize in Section~\ref{sec:conclusion}.

\section{Model set-up}\label{sec:model_setup}

\begin{figure*}[ht]
\gridline{\fig{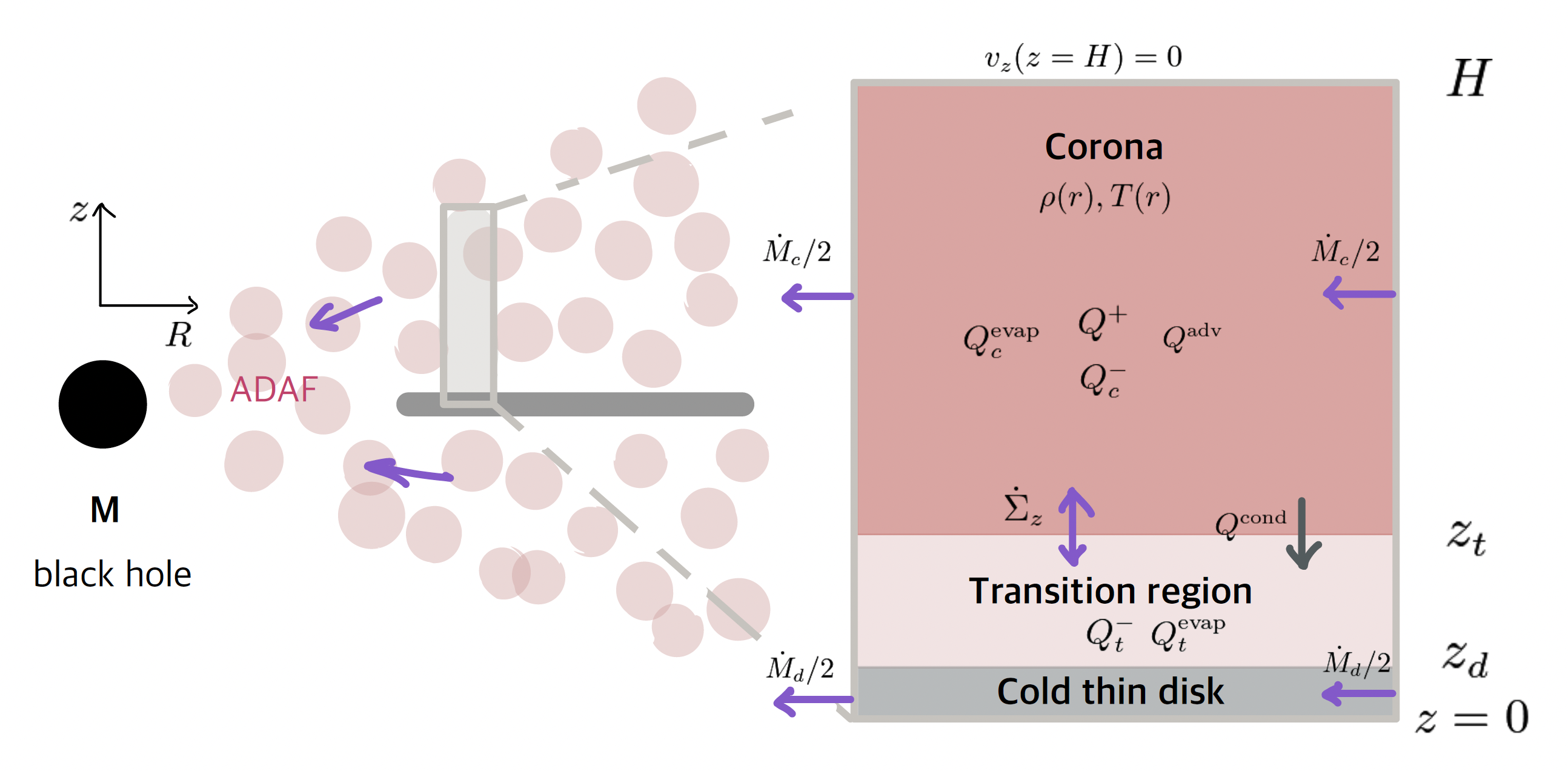}{.9\textwidth}{}
}
\caption{Schematics of the model set-up. The left half shows the black hole system in the hard state with a hot accretion flow (ADAF) at smaller radii and a cool thin disk with overlying hot corona at larger radii. The hot flow is shown as pink shaded circles and the cold disk is shown as a gray bar. The box in the right half is a zoomed-in cross-section of the thin accretion disk and corona at some radius. The disk mid-plane is at $z=0$. The boundary between the thin disk and the transition region is at $z_d$, the boundary between the transition region and the corona is at $z_t$, and the scale-height corresponding to the top of the corona is at $z=H$. In our two-zone model, the equations are height-integrated separately for the transition region ($z_d$ to $z_t$) and the corona ($z_t$ to $H$). Quantities like $Q^+$, $Q^{-}_t$, etc., are defined in Section~\ref{sec:eq_tr} and Section~\ref{sec:eq_corona}. The total radial accretion rate in the corona is $\dot{M}_c$ and the total thin disk accretion rate is $\dot{M}_d$. Each quantity is divided by $2$ in the figure because the zoomed-in box shows only the top half of the system.
The vertical mass flow rate per unit area from/to the corona is denoted as $\dot{\Sigma}_z$ and is defined in Equation~\eqref{eq:def_evaprate}. 
Mass fluxes are shown with purple arrows and energy flows with gray arrows. 
\label{fig:schematic}}
\end{figure*}

\subsection{The MM Model}\label{sec:MM_model}

We begin by briefly reviewing the basics of the MM model. The left half of Figure~\ref{fig:schematic} is a schematic of the accretion flow in the hard state. At large radii, accretion occurs via a sandwich structure in which a cool, geometrically thin equatorial disk (the gray bar) lies between two hot, geometrically thick coronal layers (shaded circles) above and below it. The cool disk evaporates fully and disappears at a certain radius (the inner edge of the gray bar), leaving the corona to continue inward as a hot accretion flow, labeled ADAF. In the thermal state, the sandwich structure extends down to the ISCO, and there is no separate ADAF zone.

The MM model deals with the vertical structure of the disk-corona region of the flow, indicated by the narrow vertical rectangle. Conservation laws are solved within this region, using appropriate boundary conditions at the bottom, where the corona meets the thin disk, and allowing for an outflowing wind at the top. The flow dynamics are simplified in a number of respects. Steady state ($\partial/\partial t=0$) and axisymmetry ($\partial/\partial\phi=0$) are assumed. The angular velocity $\Omega$ and azimuthal velocity $v_\phi$ in the corona are taken to be Keplerian,
\begin{equation}\label{eq:orbital_velocity}
    \Omega = \Omega_K \equiv  \sqrt{\frac{GM}{R^3}}, \qquad v_\phi=R\Omega_K = \sqrt{\frac{GM}{R}},
\end{equation}
where $M$ is the mass of the central BH and $R$ is the cylindrical radius. The radial velocity $v_r$ is assumed to be given by
\begin{equation}\label{eq:radial velocity}
    v_r = -\alpha\frac{c_s^2}{v_\phi}, \qquad c_s^2 \equiv \frac{P}{\rho}, 
\end{equation}
where $c_s$ is the isothermal sound speed, $P$ is the pressure, $\rho$ is the density, 
and $\alpha$ is the dimensionless viscosity parameter. There is one additional key assumption in the model regarding radial derivatives, which is best explained by considering the mass conservation equation in cylindrical coordinates,
\begin{equation}
    \mathbf{\nabla}\cdot(\rho \mathbf{v}) \equiv \frac{1}{R}\,\frac{\partial}{\partial R}(R\rho v_r) + \frac{\partial}{\partial z}(\rho v_z)=0. \label{eq:mass_cons}
\end{equation}
The MM model retains the $z$-derivative in this equation but eliminates the $R$-derivative via an ansatz,
\begin{equation}\label{eq:MM_rad_approx}
    \frac{1}{R}\,\frac{\partial}{\partial R}(R\rho v_r) \to -\frac{2}{R}\,\rho v_r,
\end{equation}
where the particular coefficient $-2$ is motivated heuristically. A similar approximation is used to eliminate the $R$-derivative in the energy equation as well. The reason for this approximation is to reduce the problem to ordinary differential equations in $z$. The complete equations are given in Appendix~\ref{sec:appendix_meyer}, where one more approximation, a convenient method of modeling the transition from cylindrical to spherical geometry for the outflowing wind, is also noted. 

\begin{figure}[ht]
\gridline{\fig{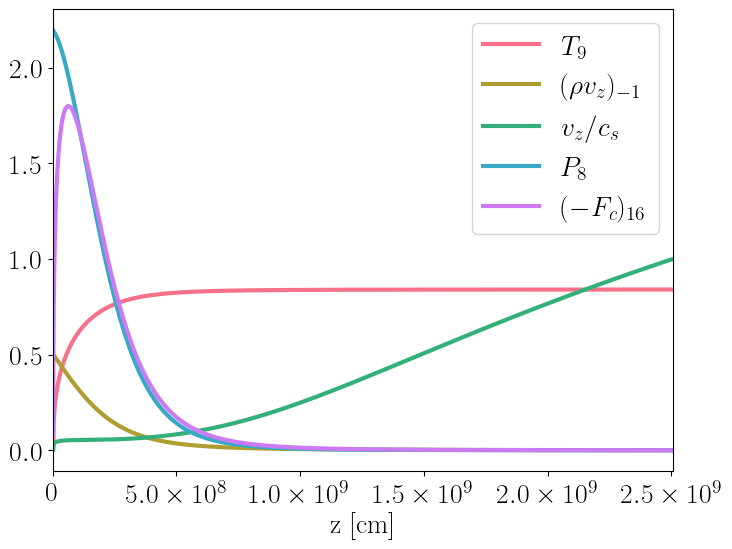}{.5\textwidth}{}
}
\caption{Numerical solution of the vertical structure of the corona at $R=10^{8.8}\,{\rm cm}$ for an accreting $6M_\odot$ black hole (the parameters have been selected to match the model shown in Figure~2 of \mlmh). Temperature $T$ is in units of $10^9 \,{\rm K}$, vertical mass flux $\rho v_z$ is in units of $10^{-1}\,{\rm g\,cm^{-2}\,s^{-1}}$, vertical velocity $v_z$ is in units of sound speed $c_s$, pressure $P$ is in units of $10^8\,{\rm dyn\,cm^{-2}}$, and conductive flux $F_c$ is in units of $10^{16}\,{\rm erg\,cm^{-2}\,s^{-1}}$. A consistent solution satisfying all boundary conditions is obtained when $\evaprate=(\rho v_z)(z=z_d)=10^{-1.30}\,{\rm g\,cm^{-2}\,s^{-1}}$ and $P(z=z_d)=10^{8.34}\,{\rm dyn\,cm^{-2}}$. Numerical calculations start from $z=z_d\ll R$ and end at the sonic height $z=z_s$, where $v_z\approx c_s$ and the conductive flux vanishes ($F_c\approx 0$). Note that the temperature rises steeply at low $z$, and becomes nearly constant (${\rm d}T/{\rm d}z\approx 0$) over a wide range of higher $z$. We identify the former as the transition region, and the latter as the corona. Also, note that the vertical mass flux $\rho v_z$ at $z=z_s$ is very small compared to the flux at the bottom ($z=z_d$). This means that mass loss in a wind is fairly weak.
\label{fig:vertical_meyer}}
\end{figure}

With the above approximations, the MM model reduces to a set of four ordinary differential equations in $z$, which correspond to the conservation of mass, vertical momentum and energy, and an expression for the conductive energy flux in terms of the temperature gradient. Figure~\ref{fig:vertical_meyer} shows a numerical solution of the equations for parameters similar to those used by \mlmh in Figure 2 of their paper. Such numerical solutions may be computed at different cylindrical radii R, and then stitched together to explore the radial structure of the corona.

\subsection{Vertically Integrated Two-Zone Model}\label{sec:Two-zone}

Accretion disk models that use height-integrated equations are useful for many applications. Such ``one-zone" models describe the disk at each radius in terms of a single average density $\Bar{\rho}$, single temperature $T$, etc. Thus, the sound speed in Equation~\eqref{eq:radial velocity} becomes $c_s^2\equiv P/\Bar{\rho}$ in the one-zone model. Instead of solving for the detailed vertical structure, one computes an effective vertical scale-height $H$ using the condition of vertical hydrostatic equilibrium, and uses this to estimate the mass surface density $\Sigma$,
\begin{equation}\label{eq:scaleheight}
H = \frac{c_s}{\Omega_K}, \qquad \Sigma = \Bar{\rho} H.
\end{equation}
Quantities like $\Bar{\rho}$, $T$, $\Sigma$ are functions of the radius and satisfy ordinary differential equations in $R$. Often one can ignore the radial derivatives and solve directly for the quantities of interest at each radius, as is done successfully in the thin accretion disk model \citep{Shakura1973}. Alternatively, one can retain the derivatives and look for analytic (usually power-law) solutions, as is done in the ADAF model \citep{Narayan_Yi_1994,Narayan1995}.

The MM disk evaporation model is not suitable for a one-zone approach since the numerical solution shown in Figure~\ref{fig:vertical_meyer} has a complicated vertical structure. This is why previous efforts to utilize a one-zone-like approach to describe re-condensation of a hot corona to an inner cool disk started with a one-zone ADAF solution and then modified it by considering the interaction with the underlying cool disk \citep{Liu2007,Meyer2007,Taam2008}. However, a two-zone model appears promising. Consider the vertical temperature profile in Figure~\ref{fig:vertical_meyer}. Within a narrow range of $z$ above the thin disk, the temperature $T$ rises sharply until it is almost equal to the coronal temperature. Above that $z$, the temperature remains almost constant over a wide range of $z$. Thus we might be able to model the system in terms of two zones: a relatively thin transition zone lying just about the thin disk, and a geometrically thick corona above that. The two-zone structure is equally obvious in the downward conductive energy flux $-F_c$. The flux starts at zero at the top of the corona and increases steadily with decreasing $z$ until it reaches a peak value close to the thin disk, before plunging down to zero. In our work, we define the $z$ coordinate corresponding to the peak of $-F_c$ as the boundary $z_t$ dividing the transition region and the corona. 

Motivated by the above considerations, we consider below vertically integrated equations where we treat the transition region and the corona as two separate zones. This is shown schematically in the right-hand side of Figure~\ref{fig:schematic}. At a given $R$ in our model, the thin disk extends vertically from $z=0$ to $z_d$, and the transition region from $z_d$ to $z_t$, where both $z_d$ and $z_t$ are small compared to $R$. The corona zone extends from $z=z_t$ to $H$, where $H$ is typically comparable to $R$.  Mass and energy flow vertically between the different zones and also radially in the corona. Note in particular that there is a radial mass accretion rate $\dot{M}_c$ in the corona, which is in addition to the mass accretion rate $\dot{M}_d$ in the thin disk.

\subsection{Vertically Integrated Continuity Equation}

Using our simplifying approximations, we integrate the mass conservation Equation~\eqref{eq:mass_cons} vertically from $z_d$ to $H$. Contrary to the MM model, which eliminates the $R$-derivative via the ansatz Equation~\eqref{eq:MM_rad_approx}, we keep the $R$-derivative but eliminate the $z$-derivative by integrating vertically.
In addition, since the numerical solution in Figure~\ref{fig:vertical_meyer} shows that the vertical mass outflow rate ($\rho v_z$) into a wind at the top of the corona is negligibly small, we assume in the height-integrated model that there is no wind: $(\rho v_z)(z=H)\approx 0$. Therefore, the height integrated version of Equation~\eqref{eq:mass_cons} becomes
\begin{equation}\label{eq:cont_eq}
     \frac{1}{R}\dv{\,}{R}\left(R\Sigma v_r \right) = \evaprate ~~ {\rm (continuity)},
\end{equation}
where 
\begin{equation}\label{eq:def_evaprate}
    \evaprate\equiv (\rho v_z)|_{z=z_d}
\end{equation} is the mass evaporation rate (if the quantity is negative, its magnitude would represent the condensation rate) from the thin disk per unit area [${\rm g\,s^{-1}\,cm^{-2}}$].

We neglect the radial mass flow in the transition region because of the small vertical thickness of this zone as well as the relatively low temperature, which suppresses the radial velocity (see Equation~\ref{eq:radial velocity}). Therefore, the entire evaporated mass flux $\evaprate$ from the thin disk travels through the transition region and reaches the corona. Effectively, this means that the continuity equation in the transition region is trivial, and Equation~\eqref{eq:cont_eq} should be viewed as being applied only to the corona zone. 

There are three unknowns at each radius $R$ in our vertically integrated model: coronal temperature $T$ (which determines $v_r$), 
surface density $\Sigma=\Bar{\rho}H$, and mass evaporation rate $\evaprate$. Equation~\eqref{eq:cont_eq} provides one ordinary differential equation to solve for these quantities. We need two more equations, which we obtain by vertically integrating the energy equation in the transition region and the corona, respectively, as we discuss next.

\subsection{Vertically Integrated Energy Equation of the Transition Region}\label{sec:eq_tr}

The full 2D partial differential equation for energy conservation in the entropy form is
\begin{equation}\label{eq:energy_eq_full}
    \rho T \mathbf{v}\cdot \mathbf{\nabla}s \equiv \rho T v_r\,\frac{\partial s}{\partial R} + \rho T v_z\frac{\partial s}{\partial z} = q^+ - q^- - \mathbf{\nabla}\cdot\mathbf{F_c},
\end{equation}
where (apart from an additive constant)
\begin{equation}\label{eq:entropy}
    s=\frac{k}{\mu}\ln{\left[\frac{T^{1/(\gamma-1)}}{\rho}\right]}
\end{equation}
is the entropy per unit mass, $k$ is the Boltzmann constant, $\mu=0.62 m_p$ is the mean molecular weight for a fully ionized gas of cosmic abundances (X=0.7 and Y=0.28), and $\gamma=5/3$ is the ratio of the specific heats\footnote{The entropy formula in Equation~\eqref{eq:entropy} is derived from the first law of thermodynamics, $T {\rm d}s = {\rm d} u+ P {\rm d}(1/\rho)$, where $u=P/((\gamma-1)\rho)$ is the internal energy per unit mass for an ideal gas, and we treat $\gamma$ as a constant. \citet{Quataert1999_entropy} describe a more careful approach to the entropy for the case when the pressure is not dominated by gas pressure alone. They present an effective adiabatic index $\gamma$ in their Equation (17) which is a weighted sum of gas and magnetic adiabatic indices.}. In the middle expression in Equation~\eqref{eq:energy_eq_full}, the first term represents the radial advection of entropy and the second term describes the vertical transport of entropy. For the quantities in the right, $q^+$ and $q^-$ are the heating rate and radiative cooling rate per unit volume, and the last term is the energy loss rate via the divergence of the conductive heat flux $F_c$. 

In the case of the transition region, we only consider the dominant energy terms. Radial advection and viscous heating are neglected because of the thinness of the transition region and the significantly cooler temperature compared to the corona. Vertical advection, the second term in the middle expression in Equation~\eqref{eq:energy_eq_full}, is important. To compute the contribution from this term, we note that
$\rho v_z=\evaprate$ is a constant in the transition region. Also, the gas is heated at constant pressure, and $T(\partial s/\partial T)_P = c_P$\footnote{The constant pressure assumption in the transition region is in part motivated by the vertical structure from \mmh. It is also an assumption used in the solar corona work of \citet{Shmeleva1973} where \mmh's fourth boundary condition originates from (in Section~\ref{sec:appendix_meyer_bc}).}. Therefore, the integrated vertical advection term, which we call the evaporative cooling term, is given by 
\begin{eqnarray}\label{eq:def_Qevap}
     Q_t^{\rm evap}(R)\equiv \int_{z_d}^{z_t}\rho v_z T \left(\frac{\partial s}{\partial z}\right) dz  = \evaprate c_P \int_{z_d}^{z_t} \frac{dT}{dz}\,dz \approx \evaprate\frac{\gamma}{\gamma-1}\frac{kT}{\mu}.
\end{eqnarray}
Here, $T$ represents the coronal temperature, $T=T(z_t)=T(H)$, and
we have assumed that the gas starts at zero temperature (the gas actually starts at the temperature $T_d$ of the thin disk, but $T_d \ll T$ and is therefore negligible). Note that we use the symbol $Q$ to represent heating/cooling rate per unit area, to distinguish it from the symbol $q$ used earlier for the rate per unit volume. Also, note that we include a subscript `t' to indicate that the evaporative cooling rate computed here refers to the transition region. We later introduce a second evaporative (or vertical advection) contribution $Q_c^{\rm evap}$ for the corona.

Considering next radiative cooling, we write the height-integrated term as
\begin{equation}
    Q^-_t(R)\equiv \int_{z_d}^{z_t} q^-\,{\rm dz}.
\end{equation}
Since the transition region spans a wide range of temperatures, cooling in this zone is very complicated and involves multiple atomic processes. Fortunately, we can use a simplifying approximation based on studies of the solar corona.
\citet{Johnston2017} computed numerical solutions of the corona and transition region in the Sun and noted that the vertically integrated radiative loss in the transition region is approximately the same as the cooling loss in the corona (see Equation~13 and Figure~3 in their paper). Assuming their result to be valid for coronae in general, we set
\begin{equation}\label{eq:rad_cooling_approx}
    Q^-_t(R)\approx Q^-_c(R)
    .
\end{equation}
To verify the validity of this assumption, we compared thee magnitudes of $Q^-_t$ and $Q^-_c$ by vertically integrating the numerical solution of the MM model which uses a realistic cooling function. We confirmed that the two terms differ by only a factor of a few ($\sim 2-3$ at worst) over radii in the range $r \sim 10^{3.5}-10^{4.5}$ where the MM model is most reliable. (As described in Section~\ref{sec:appendix_meyer}, at radii $r<10^3$, the MM model does not self-consistently model the condensing corona, and at radii $r>10^5$, the coronal temperature approaches the lower temperature boundary condition of $10^{6.5}\,{\rm K}$ used by the MM model.)
We note further that our assumption in Equation~\eqref{eq:rad_cooling_approx}, which is inspired by numerical studies of the solar corona \citep{Johnston2017}, replaces a different approximation in the MM model, which also originates from solar corona research \citep{Shmeleva1973}, where the fourth boundary condition in Section~\ref{sec:appendix_meyer_bc} relates the conductive flux and pressure at the bottom of the transition region.
We postpone discussion of the coronal cooling term $Q_c^-$ to Section~\ref{sec:eq_corona}.

The last term in the energy equation of the transition region is the conductive energy flux flowing into this region from the corona. As discussed earlier, the downward conductive flux $-F_c$ reaches its maximum value at the boundary between the corona and the transition region. Thus $-F_c(z_t)$ is the net conductive heating rate of the transition region per unit area. We therefore write
\begin{eqnarray}\label{eq:def_Qcond}
    Q^{\rm cond}(R)\equiv -\int_{z_d}^{z_t} \pdv{F_c}{z}\,{dz}\approx -F_c(z_t), 
\end{eqnarray}
where we have made the reasonable assumption that $F_c(z_d)=0$, i.e., the energy loss via conduction from the transition region into the thin disk at $z=z_d$ is negligible.

Putting all the terms together, the height-integrated energy equation for the transition region is
\begin{eqnarray}\label{eq:energy_tr}
      Q^{\rm evap}_t(R) + Q^-_t(R) = Q^{\rm cond}(R) ~~{\rm (transition~region)}.
\end{eqnarray}
In words, the conductive heat flux flowing in from the corona $Q^{\rm cond}$ provides energy to the transition region, and this energy is used partly to supply the evaporation energy $Q_t^{\rm evap}$ and the rest is lost via radiative cooling $Q^-_t$. Equation~\eqref{eq:energy_tr} is a second condition which the height-integrated model must satisfy and which we can use to solve for the three unknowns in the problem. Note that, compared to Equation~\eqref{eq:cont_eq}, which is a differential equation, here we have an an algebraic equation with no radial derivatives.

\subsection{Vertically Integrated Energy Equation of the Corona}\label{sec:eq_corona}

When considering the corona, we keep all the terms in Equation~\eqref{eq:energy_eq_full} and integrate from $z=z_t$ to $H$.
The first term is the radial advective energy term, which becomes after vertical integration,
\begin{eqnarray}
    Q^{\rm adv}(R) \equiv \int_{z_t}^H \rho T v_r \pdv{s}{R}\,dz
    \approx\Sigma v_r \frac{kT}{\mu} \left[\frac{1}{(\gamma-1)T}\dv{T}{R}-\frac{1}{\Sigma}\dv{\Sigma}{R}\right].
\end{eqnarray}
Note that quantities like $\Sigma$ and $H$ correspond to one side of the corona from the mid-plane. The temperature $T$ is assumed to be independent of $z$ in the corona, as described in Section~\ref{sec:Two-zone}.

The second term in the middle of Equation~\eqref{eq:energy_eq_full} describes vertical energy advection. This is another component of evaporation energy, which we call $Q_c^{\rm evap}$. As the gas rises in the corona, it moves at constant temperature $T$ but gains entropy because its density declines steadily with increasing $z$. Using the entropy expression in Equation~\eqref{eq:entropy} and assuming a constant $T$, we estimate the evaporative energy per unit area to be
\begin{equation}\label{eq:Qevapc}
     Q_c^{\rm evap}(R) \equiv \int_{z_t}^H \rho T v_z \pdv{s}{z}\,dz =-\int_{z_t}^H \rho v_z \frac{kT}{\mu} \dv{\ln{\rho}}{z} \,dz \approx 
     \zeta\evaprate\frac{\gamma}{\gamma-1}\frac{kT}{\mu}.
\end{equation}
It is hard to estimate the above integral precisely, but we expect its value to be similar to the corresponding term in the transition region, $Q^{\rm evap}_t$.
For convenience, and in the spirit of our toy model, we simply assume that the evaporation energy $Q^{\rm evap}_c$ in the corona is equal to $Q^{\rm evap}_t$ in the transition region multiplied by a factor of order of unity $\zeta$.
We thereby obtain the approximate expression given at the right of Equation~\eqref{eq:Qevapc}. By vertically integrating the numerical solutions of the MM model, we find that $\zeta$ lies in the range $0.6-0.7$ for radii in the range $r\sim 10^{3.5}-10^{4.5}$. Moreover, when we keep $\zeta$ and propagate it through the analytic solutions discussed later, the dependence on $\zeta$ is weak, e.g., $T, ~\Sigma \propto (2+5\zeta)^{-1/3}$ for the direct magnetic heating model discussed in Section~\ref{sec:analytic_sols_mag}. Therefore, we assume $\zeta\sim1$ for simplicity.
With this approximation, we end up with $Q^{\rm evap}_c = Q^{\rm evap}_t$, hence we hereafter use $Q^{\rm evap}$ without a subscript if referring to either term.

Considering next the heating term in Equation~\eqref{eq:energy_eq_full}, vertical integration gives
\begin{equation}
    Q^+(R) \equiv \int_{z_t}^H q^+ \,dz\approx q^+ H.
\end{equation}
In the present section, and in Sections~\ref{sec:radial_analytic} - \ref{sec:truncation_radius}, we assume that the only source of heating in the corona is viscous dissipation (we consider an additional source of heating in Section~\ref{sec:mag}). For this case, we have
\begin{equation}\label{eq:def_Qplus_visc}
Q^+=Q^+_{\rm visc}\approx \frac{3}{2}\alpha P\Omega_K H.
\end{equation}
We write the pressure as
 \begin{equation}
     P=\pcorr P_{\rm gas}=\pcorr \frac{\Bar{\rho} kT}{\mu},
 \end{equation}
where the coefficient $\pcorr$ describes any additional contribution to the pressure over and above gas pressure $P_{\rm gas}$, and $\Bar{\rho}\equiv \Sigma /H$ is the average density of the corona. For example, \citet{Meyer2002} consider the case when magnetic pressure is not negligible. In terms of the plasma-$\beta$, they write
\begin{equation}\label{eq:pcorr_to_beta}
\pcorr=1+\frac{1}{\beta}, \qquad \beta\equiv \frac{P_{\rm gas}}{P_{\rm mag}}= \frac{\Bar{\rho} kT/\mu}{B^2/8\pi}.
\end{equation}

Considering next radiative cooling, the corresponding height-integrated term in the corona is
\begin{equation}
    Q^-_c(R)\equiv \int_{z_t}^H n_e n\Lambda(T) \,{\rm dz}\approx n_e n\Lambda(T) H=Q^-_{\rm Brem}(R),\label{eq:Qminus}
\end{equation}
where $n_e=(1+X)\Bar{\rho}/(2 m_p)$ and $n=(X+Y/4)\Bar{\rho}/m_p$ are the number densities of electrons and ions respectively (see below Equation~\eqref{eq:entropy} for the values of $X$ and $Y$ used), and $\Lambda(T)$ is the cooling function of the gas. In the present study we use
\begin{equation}
\Lambda(T)=fT^{1/2}, \qquad f\approx 2.8\times 10^{-27}\,{\rm erg\,cm^3\,s^{-1}\,K^{-1/2}},
\end{equation}
where the coefficient corresponds to the \brem curve in Figure~1 of \citet{Raymond1976}. This estimate of the coefficient agrees with Equation (34.3) in \citet{Draine2011} to within a factor of a few. We use \brem cooling instead of a more detailed cooling curve.
However, as we argue in Appendix~\ref{sec:appendix_cooling}, this approximation is actually quite reasonable. In view of the approximation in Equation~\eqref{eq:rad_cooling_approx}, the radiative cooling rate in the corona and the transition region are the same, and we refer to the corresponding rate as $Q^-_{\rm Brem}$ referring to height-integrated \brem cooling rate.

We note that Compton cooling is likely to be important close to the black hole. We discuss this issue in Section~\ref{sec:compton_cooling} and we include an extra Compton cooling term in Section~\ref{sec:analytic_sols_mag} when we discuss a generalized model where the corona is additionally heated by direct coronal heating.

The conduction term in Equation~\eqref{eq:energy_eq_full} is a divergence of the conductive heat flux. Therefore, when we vertically integrate, the result is the difference between the fluxes at the two boundaries. The conductive flux at the top of the corona is negligible (this is in fact one of the boundary conditions in the MM model). Therefore, we obtain the same formula for conduction as $Q^{\rm cond}$ in Equation~\eqref{eq:def_Qcond}
\begin{eqnarray}
    \int_{z_t}^H \pdv{F_c}{z}\,{dz}\approx -F_c (z_t).
\end{eqnarray}
In an unmagnetized plasma, the vertical heat flux is given by
\begin{equation}\label{eq:def_cond_flux_classical}
    F_c =-\kappa_0 T^{5/2} \pdv{T}{z} = -\frac{2}{7}\kappa_0 \pdv{(T^{7/2})}{z}, \qquad \kappa_0 =10^{-6} \,{\rm g\, cm\,s^{-3}\,K^{-7/2}}.
\end{equation}
Allowing for a possible modest suppression of conduction, we thus write
\begin{eqnarray}
    Q^{\rm cond}(R) \approx -F_c (z_t)\approx \frac{2}{7}\Tilde{\kappa} \kappa_0 \frac{T^{7/2}}{H} , 
\end{eqnarray}
where $\Tilde{\kappa} \leq 1$ is a numerical factor. An organized frozen-in field strongly suppresses cross-field conduction in a plasma. Since we expect the magnetic field in the corona to emerge from the underlying thin disk in a roughly vertical direction and to be swept back azimuthally, we do not expect much radial conduction, which is why we ignored the radial part of $\mathbf{\nabla}\cdot\mathbf{F_c}$. For the vertical direction of interest to us, we expect classical conduction to operate if the field is uniform, but the flux could be reduced if there are non-uniformities in the field.
\citet{Narayan2001} showed that a turbulent magnetic field can reduce conductivity from the classical value by a factor of a few, $1/5\lesssim \kappacorr \lesssim 1$. This is a reasonable range of values to consider in the present work. 

Combining all the terms, the vertically integrated energy equation for the corona is
\begin{eqnarray}\label{eq:energy_corona}
    \Sigma v_r \frac{kT}{\mu} \left[\frac{1}{(\gamma-1)T}\dv{T}{R}-\frac{1}{\Sigma}\dv{\Sigma}{R}\right] = Q^+(R) - Q^{\rm evap}_c(R) - Q^-_{\rm Brem}(R) - Q^{\rm cond}(R) ~~{\rm (corona)},
\end{eqnarray}
where the left-hand side corresponds to $Q^{\rm adv}(R)$. Note that $Q^{\rm cond}$ is an energy flux that flows out of the corona into the transition region. It is a cooling term for the corona, but becomes a heating term for the transition region.

We now have three equations, viz., Equations~\eqref{eq:cont_eq}, \eqref{eq:energy_tr}, \eqref{eq:energy_corona}, for the three unknowns at each radius, viz., $T(R)$, $\Sigma(R)$, and $\dot{\Sigma}_z(R)$. Two of the equations are first-order differential equations in $R$, and the third (Equation~\ref{eq:energy_tr}) is an algebraic relation. The problem is thus reduced to a second-order differential equation in $R$.

\subsection{Momentum equations}\label{sec:eq_momentum}

The three equations we derived above come from the continuity equation and the energy equation. While we have not explicitly mentioned the 
momentum equations, they are effectively used in the model. Following \mlmh, the azimuthal velocity is approximated to be Keplerian (Equation~\ref{eq:orbital_velocity}) using the $R$-component of the momentum equation, and the radial velocity formula Equation~\eqref{eq:radial velocity} is obtained from the $\phi$-component of the momentum equation. Finally, the $z$-component of the equation is written as a condition for vertical hydrostatic equilibrium, whereby the formula for the scale-height $H$ (Equation~\ref{eq:scaleheight}) is obtained.

\subsection{Assumptions in the Current Model}\label{sec:assumptions}

Here we list the assumptions used for constructing vertically integrated equations in the model. 
\begin{itemize}
    \item In the transition region, we assume that the pressure is nearly constant in order to obtain the formula for the evaporative cooling term $Q^{\rm evap}_t$ in Equation~\eqref{eq:def_Qevap}.
    
    \item Due to the lower temperature of the transition region compared to the corona and the vertical thinness of this region, we neglect the radial advective term $Q^{\rm adv}$ and the heating term $Q^+$.
    
    \item We assume that the vertically integrated radiative cooling in the corona and the transition region are equal ($Q^-_t\approx Q^-_c\approx Q^-_{\rm Brem}$), except when we include Compton cooling as discussed in Section~\ref{sec:analytic_sols_mag}.
    
    \item For simplicity of the model, we also assume $Q^{\rm evap}_t \approx Q^{\rm evap}_c$. From vertical solutions of \mmh, we confirm that this assumption is good.
    
    \item In the corona, we assume that the temperature is nearly independent of $z$, which is reasonable as can be seen in Figure~\ref{fig:vertical_meyer}.
    
    \item We neglect conductive flux at the bottom of the transition region $F_c(z_d)\approx 0$ to obtain Equation~\eqref{eq:def_Qcond}.
    
    \item We assume that ions and electrons are coupled such that the plasma is well described by a single temperature for both species.
    
    \item We neglect coronal outflow, i.e., we set $v_z(z=H)\approx 0$.
\end{itemize}

\section{Radial Solutions of the Vertically-Integrated Corona Model}\label{sec:radial_analytic}

The three equations describing our model, Equations \eqref{eq:cont_eq}, \eqref{eq:energy_tr}, and \eqref{eq:energy_corona}, involve several terms and cannot be solved trivially. One approach is to focus only on the dominant terms, judiciously selected, and to look for analytic power-law solutions of the form
\begin{equation}\label{eq:powerlaw}
    T(R)\propto R^p,\qquad \Sigma(R) \propto R^q.
\end{equation}
Alternatively, we could retain all the terms and solve the equations numerically. We discuss both approaches in the following subsections and present some results. In the plots, we use $M=10M_\odot$, $\alpha=0.3$, $\kappacorr=1$, $\pcorr=1$, unless otherwise stated. However, as we show, most of the results are independent of $M$ when we scale quantities suitably.

\subsection{Analytical Solutions: Evaporation-Dominated and Radiative Cooling-Dominated Regimes}\label{sec:two_regimes}

We begin by considering the energy balance condition, Equation~\eqref{eq:energy_tr}, in the transition region, which is an algebraic relation between a single heating term (conduction) and two cooling terms (evaporation and radiative cooling). It is natural to think that, depending on conditions, one of the cooling terms will dominate over the other. We can thus imagine two distinct regimes: (i) an evaporation-dominated regime where $Q^{-}_{\rm Brem} \ll Q^{\rm evap} \approx Q^{\rm cond}$, and (ii) a radiative cooling-dominated regime where $Q^{\rm evap} \ll Q^{-}_{\rm Brem} \approx Q^{\rm cond}$.

We first consider the evaporation-dominated regime,
\begin{equation}
    Q^{\rm cond}(R)\approx Q^{\rm evap}(R) \qquad {\rm (evaporation~dominated~regime)}
    ,
\end{equation}
where almost all the conductive heat flux from the corona is used to evaporate gas from the thin disk into the corona, and there is very little radiative cooling. Neglecting the cooling term in the equations, it is then possible to obtain a power-law solution with indices $p=-1$, $q=-2$, i.e., $T\propto R^{-1}$, $\Sigma\propto R^{-2}$. The analytic solution is written explicitly in Appendix Section~\ref{sec:appendix_viscous}, where the radius $R$ is scaled to the Schwarzschild radius $R_S$ as follows, 
\begin{equation}
    r\equiv \frac{R}{R_S}=\frac{R}{2 GM/c^2}.
\end{equation}
From the analytical solution, we find that the two cooling terms scale as follows:
\begin{equation}\nonumber
    \left|\frac{Q^{\rm evap}}{Q^-_{\rm Brem}}\right|\propto T^{2} \Sigma^{-1} R\propto R.
\end{equation}
We see that the evaporation term $Q^{\rm evap}$ dominates at large $R$. Hence, the evaporation-dominated regime corresponds to radii far from the central black hole.

In the opposite case, when radiative cooling dominates over evaporative cooling,
\begin{equation}
    Q^{\rm cond}(R)\approx Q^{-}_{\rm Brem}(R)\qquad {\rm (cooling~dominated~regime)}
    ,
\end{equation}
we can repeat the same analysis, but this time neglecting the evaporation term $Q^{\rm evap}_t$. We find that we must also neglect the radial advection term $Q^{\rm adv}$ in the corona (we discuss the validity of this approximation in Section~\ref{sec:energetics}). We then obtain a power-law solution with $p=0$ and $q=0$, corresponding to $T\propto R^0={\rm constant}$, $\Sigma\propto R^0={\rm constant}$.  
Once again, comparing the two competing cooling terms in this second regime, we find
\begin{equation}\nonumber
    \left|\frac{Q^{\rm evap}}{Q^-_{\rm Brem}}\right| \propto T^{2} \Sigma^{-1} R\propto R,
\end{equation}
which is the same as before. It is clear that cooling dominates at small radii, i.e., closer to the black hole. 

\begin{figure*}[ht]
\gridline{\fig{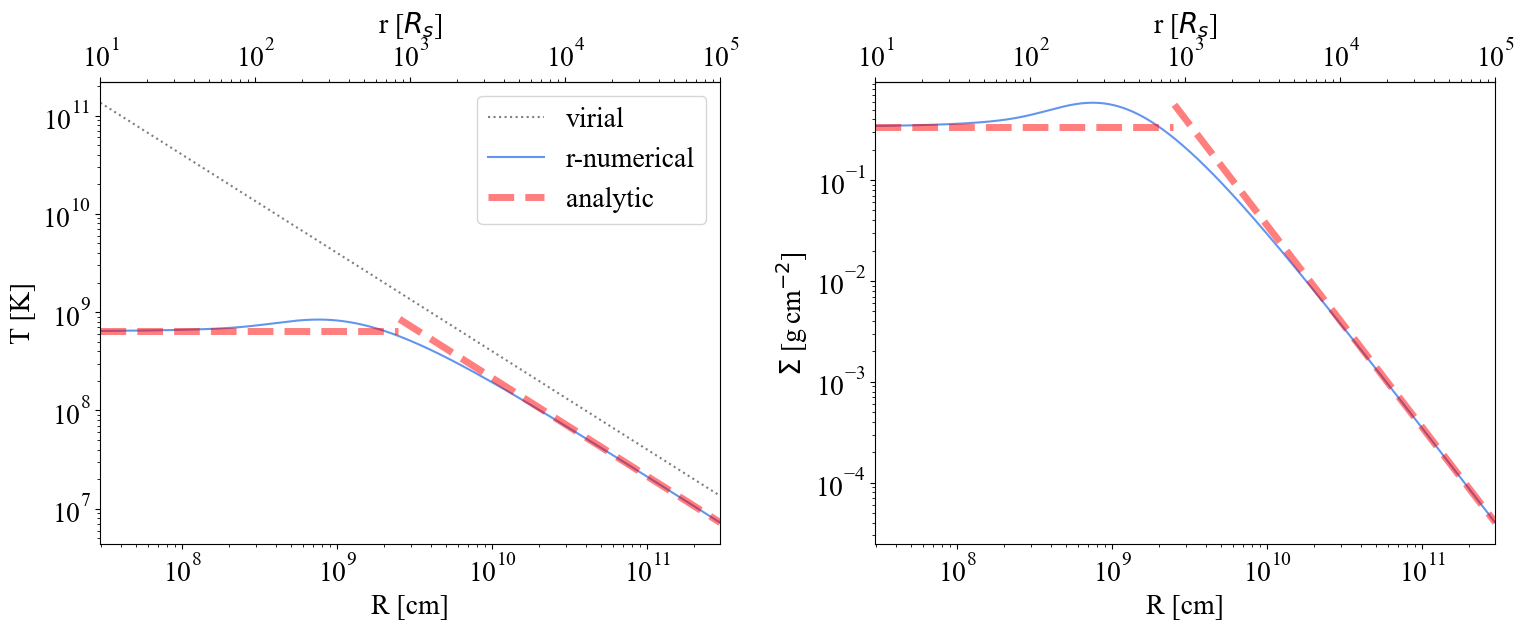}{.9\textwidth}{}
}
\caption{Analytical (Section~\ref{sec:two_regimes}) and numerical (Section~\ref{sec:numerical_sol}) solutions for the corona temperature $T$ \emph{(left)} and surface density $\Sigma$ \emph{(right)} for the vertically integrated corona model. The solutions are independent of the BH mass when expressed in terms of the Schwarzschild-scaled cylindrical radius $r$ (upper horizontal axis). The physical cylindrical radius $R$ (lower horizontal axis) corresponds to the specific case of a $10M_\odot$ BH. The continuous blue solid lines show the numerical solution, and the red dashed lines show the analytical solutions corresponding to the evaporation-dominated regime (large $r$) and cooling-dominated regime (small $r$), respectively. 
The black dotted line in the left panel corresponds to the virial temperature $T_{\rm vir}$.
\label{fig:T_rho_viscous}}
\end{figure*}

Using the analytic solutions for $T$ and $\Sigma$ in the two regimes, we can estimate the break radius $r_b \equiv R_b/R_S$ where the two zones meet. This can be done in two ways. First, we could look at the analytical solution for the evaporation-dominated regime, where $Q^-_{\rm Brem}$ is neglected, compute $Q^-_{\rm Brem}$ after the fact from the derived solution, and determine the radius at which $Q^-_{\rm Brem}$ becomes equal to $Q^{\rm evap}$. This is the radius at which the initial assumption $Q^-_{\rm Brem} \ll Q^{\rm evap}$ breaks down. Similarly, we could do the converse, using the cooling-dominated analytical solution. These two independent calculations give
\begin{equation}
r_b=
    \begin{cases}
    10^{3.02}\,\kappacorr\,\newalpha^{-2}\,\pcorr^{-3} &({\rm inner~cooling~dominated~solution}),\\
        
    10^{2.99}\,\kappacorr\,\newalpha^{-2}\,\pcorr^{-3} & ({\rm outer~evaporation~dominated~solution}),
    
    \end{cases}
\end{equation}
where $\newalpha\equiv \alpha/0.3$.
Note the identical scalings of the two results with respect to the parameters $\Tilde{\kappa}$, $\newalpha$ and $\Tilde{p}$, and the nearly identical coefficients. The close agreement indicates that the two radial zones we have identified are physically well-motivated. 
Hereafter, we define the break radius between the two regimes to be
\begin{equation}\label{eq:r_break}
    r_b \equiv 10^3\,\kappacorr\, \newalpha^{-2}\,\pcorr^{-3}.
\end{equation}
Note the interesting fact that $r_b$ is independent of the black hole mass $M$: the break between the cooling-dominated and evaporation-dominated regimes is located at the same Eddington-scaled radius for black holes of any mass.

We now write the solutions for the coronal temperature $T$, scaling the radius by $r_b$ (note that the solutions written in Appendix Section~\ref{sec:appendix_viscous} do not scale by $r_b$). This gives for the two regimes
\begin{equation}\label{eq:T_visc_sol}
T=
    \begin{cases}
    6.4\times10^8 \,\kappacorr^{-1}\,\newalpha^2\,\pcorr^3\,{\rm K} &(r<r_b),\\
    7.2\times 10^8\,\kappacorr^{-1}\,\newalpha^2\,\pcorr^{3}\,\left(\frac{r}{r_b}\right)^{-1}\,{\rm K} & (r>r_b),
    \end{cases}
\end{equation}
We immediately see that, in Schwarzschild units, the temperature is mass-independent, meaning that the coronal states are the same for all BHs from stellar-mass BHs to SMBHs. In the left panel in Figure~\ref{fig:T_rho_viscous}, the two analytic solutions for $T$ are shown as red dashed lines. Also shown for comparison is the virial temperature,
\begin{equation}
    T_{\rm vir}=\frac{\gamma-1}{\gamma}\frac{\mu GM}{kR}.
    \end{equation}
We see that the coronal temperature predicted by the model is subvirial over the entire range of radii, meaning that the corona is gravitationally well bound. This further supports our choice to ignore any wind escaping from the top of the corona.

The analytic solutions for the surface density are
\begin{equation}\label{eq:Sigma_visc_sol}
\Sigma=
    \begin{cases}
        0.34\,\kappacorr^{-1}\,\newalpha^3\,\pcorr^{\frac{9}{2}}\,{\rm g\,cm^{-2}} &(r<r_b),\\
    0.40\,\kappacorr^{-1}\,\newalpha^{3}\,\pcorr^{\frac{9}{2}}\, \left(\frac{r}{r_b}\right)^{-2}\,{\rm g\,cm^{-2}} & (r>r_b),
    \end{cases}
\end{equation}
which are again mass-independent. These solutions are shown in the right panel of Figure~\ref{fig:T_rho_viscous}. However, since at a given $r$ the scale-height $H$ is proportional to BH mass, the average coronal density $\Bar{\rho}=\Sigma/H$ does show a mass dependence,
\begin{equation}\label{eq:rho_visc_sol}
\Bar{\rho}=
    \begin{cases}
        2.6\times 10^{-9}\, \kappacorr^{-2}\,\newalpha^5\,m^{-1}\,\pcorr^7\,\left(\frac{r}{r_b}\right)^{-\frac{3}{2}}\,{\rm g\,cm^{-3}} &(r<r_b),\\
        3.0\times10^{-9}\, \kappacorr^{-2}\,\newalpha^{5} \,m^{-1}\,\pcorr^7\,\left(\frac{r}{r_b}\right)^{-3} \,{\rm g\,cm^{-3}} & (r>r_b),
    \end{cases}
\end{equation}
where $m\equiv M/M_\odot$ is the dimensionless BH mass in solar units. 

\subsection{Radial numerical solutions}\label{sec:numerical_sol}
For comparison with the analytical solutions, we compute numerical solutions by solving the full set of equations without neglecting any terms. As a first step, we solve for $\evaprate$ from Equation~\eqref{eq:energy_tr} (energy equation for the transition region), which is the only algebraic equation. Then, we substitute this in the other two equations: continuity Equation~\eqref{eq:cont_eq} and the energy equation for the corona Equation~\eqref{eq:energy_corona}. After substitution, we are left with two coupled differential equations,
\begin{eqnarray}\label{eq:explicit_eq_cont}
    \frac{1}{R}\dv{\,}{R}\left(R\Sigma v_r\right) &=&\frac{\gamma-1}{\gamma}\frac{\mu}{kT}\left(- n_e n\Lambda(T) H +\frac{2}{7}\Tilde{\kappa} \kappa_0 T^{7/2}/H\right) ,
    \\\label{eq:explicit_eq_c}
    \Sigma v_r \frac{kT}{\mu} \left[\frac{1}{(\gamma-1)T}\dv{T}{R}-\frac{1}{\Sigma}\dv{\Sigma}{R}\right] &=& \frac{3}{2}\alpha P \Omega_K H - \frac{4}{7}\Tilde{\kappa} \kappa_0 T^{7/2}/H,
\end{eqnarray}
which we solve numerically\footnote{Equation~\eqref{eq:explicit_eq_c} has the unusual form, $Q^{\rm adv} = Q^+ - 2Q^{\rm cond}$, but this is an accidental consequence of our approximations. The energy equation of the corona is $Q^{\rm adv} = Q^+ - Q_c^{\rm evap} - Q_c^- - Q^{\rm cond}$, while that of the transition region has the simpler form, $0 =  - Q_t^{\rm evap} - Q_t^- + Q^{\rm cond}$. Adding the two equations, we obtain the reasonable-looking equation, $Q^{\rm adv} = Q^+ - [Q_c^{\rm evap} + Q_t^{\rm evap}] - [Q_c^- + Q_t^-]$. However, if we instead subtract the two equations and further substitute our approximations, $Q_c^{\rm evap} \approx Q_t^{\rm evap}$ (for $\zeta=1$) and $Q_t^- \approx Q_c^- = Q_{\rm Brem}^-$, we end up with Equation~\eqref{eq:explicit_eq_c}.}. Note that $(R\Sigma v_r)$ in the left-hand side of Equation~\eqref{eq:explicit_eq_cont} is a function of both $T$ (via $v_r$) and $\Sigma$. Therefore, both Equations~\eqref{eq:explicit_eq_cont} and \eqref{eq:explicit_eq_c} contain radial derivatives of $T$ and $\Sigma$ on the left-hand side. These first-order differential equations are solved by starting with initial values of temperature $T$ and surface density $\Sigma$ at a large radius well into the outer evaporation-dominated regime. For the specific example shown in Figure~\ref{fig:T_rho_viscous}, we chose the initial values to correspond to the analytic solutions Equations~\eqref{eq:T_visc_sol} and \eqref{eq:rho_visc_sol}. However, the numerical solution is actually robust even when the initial values are changed by up to a factor of 10; in these cases, with decreasing $r$, the solution quickly adjusts and asymptotes to the solution shown in Figure~\ref{fig:T_rho_viscous}. The numerical calculation is however sensitive to the direction of integration. Sometimes, we have found that integrating numerically from small to large radii fails, but integrating in the other direction is always stable.

Figure~\ref{fig:T_rho_viscous} shows the numerical solution corresponding to the same parameters used for the analytical solutions. 
The numerical solution shows a break in slope at around $r\sim 10^3$, roughly where the analytical solution predicts the break $r_b$ between the evaporation-dominated and cooling-dominated regimes should be located (Equation~\ref{eq:r_break}). Also, the numerical solution agrees perfectly with the two analytical solutions, both in slope and normalization, at radii away from the break. 

\subsection{Energetics}\label{sec:energetics}

\begin{figure}[ht]
\gridline{\fig{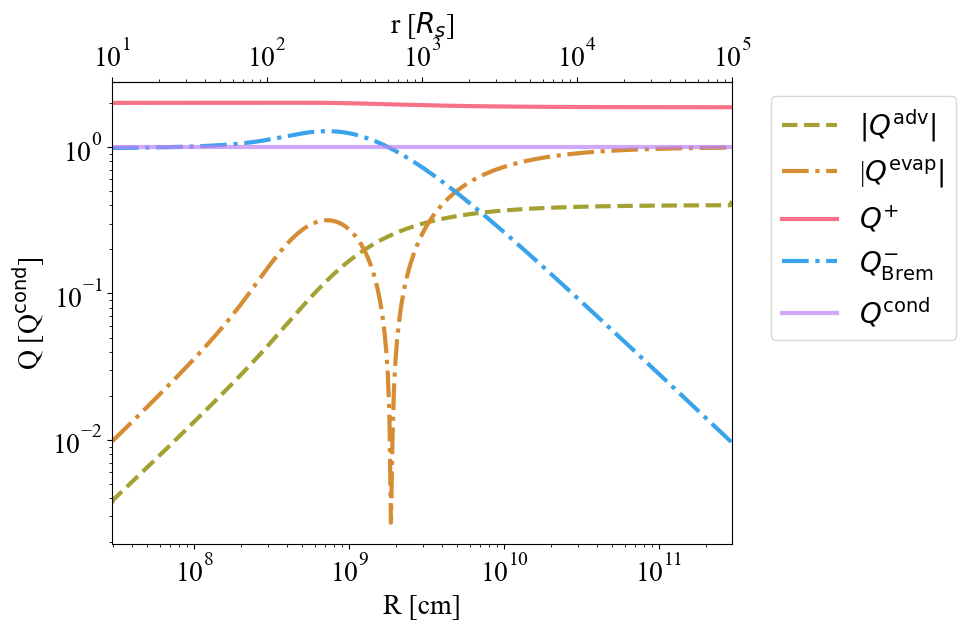}{.7\textwidth}{}
}
\caption{Energy terms normalized by the conduction term $Q^{\rm cond}$ from the radial numerical solution. Since we assume $\zeta\sim 1$ in Equation~\eqref{eq:Qevapc}, we use $Q^{\rm evap}$ without a subscript to refer to either $Q^{\rm evap}_c$ or $Q^{\rm evap}_t$. In the inner radii, the dominant energy terms are viscous heating $Q^{+}$, radiative cooling $Q^-_{\rm Brem}$, and conduction $Q^{\rm cond}$. The advection term $Q^{\rm adv}$ and evaporation term $Q^{\rm evap}$ are negligible. In the outer radii, leaving aside $Q^-_{\rm Brem}$, all the other terms, namely,  $Q^+$, $Q^{\rm cond}$, $Q^{\rm evap}$, and $Q^{\rm adv}$, are important.
}
\label{fig:energy_viscous}
\end{figure}

Here we investigate the contribution of each of the individual energy terms in the vertically integrated energy equations to obtain a better understanding of the physics of the system.
Figure~\ref{fig:energy_viscous} shows how each energy term, normalized by the conduction energy $Q^{\rm cond}$, varies with radius in the numerical solution discussed in Section~\ref{sec:numerical_sol} (shown in Figure~\ref{fig:T_rho_viscous}).
It is evident that the system has two very distinct regimes as a function of radius. At radii $r>10^3$,
all the terms are roughly comparable in magnitude, with the sole exception of the radiative cooling term $Q^{-}_{\rm Brem}$, which is negligible. On the other hand, at radii $r<10^3$, the radiative cooling term $Q^{-}_{\rm Brem}$ quickly becomes one of the dominant terms, whereas both the radial advection term $Q^{\rm adv}$ and the evaporative term $Q^{\rm evap}$ become negligible. These trends are consistent with the approximations we used to obtain the two limiting analytical solutions described in Section~\ref{sec:two_regimes}.

Interestingly, with decreasing $r$, we see in Figure~\ref{fig:energy_viscous} that $Q^-_{\rm Brem}$ first overshoots before settling down to its asymptotic solution. The overshoot coincides with a region where $Q^{\rm evap}$ changes sign, which signifies that there is a major change in the nature of the solution. Indeed, as we show in Section~\ref{sec:interaction_disk_corona}, the corona makes a transition at this point from evaporation to condensation, which is a major change in physics. This may explain in part why, with decreasing $r$ in Figure~\ref{fig:T_rho_viscous}, the numerical solution requires a factor of several in radius below the break before it settles down to the new power-law solution at small $r$.

\section{Comparison of the vertically-integrated model with the MM model}\label{sec:compare_to_meyer}

The major differences between our model and the MM model are 1) the assumption in our model that there is no outflowing wind at the top of the corona, and 2) the manner in which we treat partial derivatives in the conservation laws. Each of these differences is discussed in the following subsections.

\subsection{No-wind Approximation}\label{sec:nowind_approx}

The MM model considers disk-corona systems characterized by an outflowing wind at the top of the corona. This feature is reflected in their model as additional terms in the continuity Equation~\eqref{eq:meyer_cont_eq} and energy Equation~\eqref{eq:meyer_energy_eq}, and also via boundary conditions at the top (see Section~\ref{sec:appendix_meyer_bc}). 
As described in Section~\ref{sec:model_setup}, our model assumes that there is no wind escaping from the top of the corona, i.e., $\rho v_z=0$ at $z=H$. We test the validity of this approximation by comparing numerical solutions as a function of $z$, similar to the work of \mmh, both with and without a wind.

\begin{figure}[ht]
\gridline{\fig{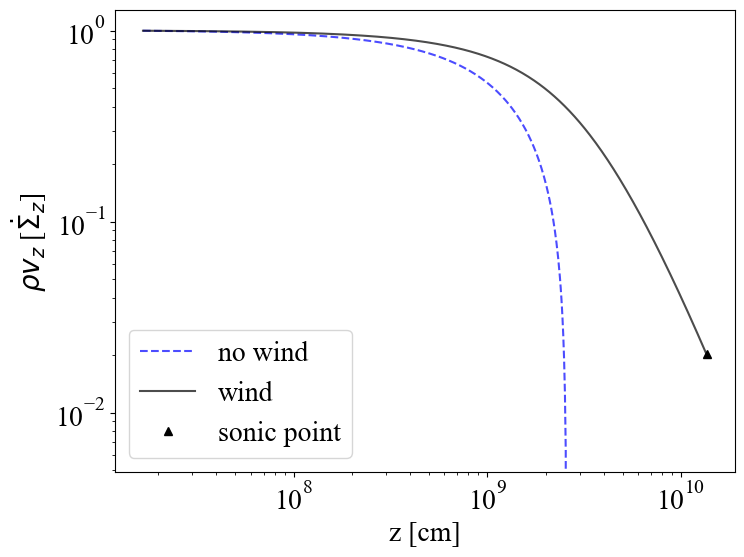}{.48\textwidth}{}
}
\caption{Vertical mass flux $\rho v_z$ as a function of height $z$, normalized by the vertical mass flux from the thin disk $\evaprate$. The calculations are for a cylindrical radius $R=10^{9.7}\,{\rm cm}=10^{3.5}R_S$ for a $6M_\odot$ BH. The black solid line shows the wind solution and the blue dashed line shows the no-wind solution. The black triangle indicates where the vertical velocity in the wind model is equal to the sound speed (sonic point).
\label{fig:mdot_vertical}}
\end{figure}

In Figure~\ref{fig:mdot_vertical}, the vertical flux of matter as a function of $z$ at a fixed $R$ is shown for both models. The ``wind" solution is calculated using the same equations and boundary conditions as in \mmh, which are briefly summarized in Section~\ref{sec:appendix_meyer_eq} and Section~\ref{sec:appendix_meyer_bc}. This solution passes through a sonic point at the top of the corona. For the ``no wind" solution, we solve a similar set of equations, but modified so as to be consistent with the assumptions of our two-zone model. Specifically, we neglect the wind terms in the equations, and we modify the boundary condition such that the vertical speed goes to zero when $z=H$ (see Section~\ref{sec:appendix_vertical_no_wind} for details).
Figure~\ref{fig:mdot_vertical} shows that the two solutions are very similar at low values of $z$. They start to diverge at larger values of $z$ near the top of the corona, but the density is low by this point. Note in particular that, at the sonic point $v_z\approx c_s$, in the wind solution, the escaping vertical mass flux is only 2.0\% of the initial mass flux $\dot{\Sigma}_z$ that flows into the corona at the bottom. Setting the mass outflow to zero in our model is unlikely to have a serious effect.


\subsection{Mass Exchange between the Thin Disk and the Corona}\label{sec:interaction_disk_corona}
As explained in Section~\ref{sec:MM_model}, the MM model simplifies the problem by eliminating radial derivatives via the ansatz given in Equation~\eqref{eq:MM_rad_approx}.
Here we examine what effect this approximation has on the solution.

Let us vertically integrate the mass conservation Equation~\eqref{eq:mass_cons}. Ignoring mass loss in a wind (including it will not change anything), and applying ansatz Equation~\eqref{eq:MM_rad_approx}, we obtain
\begin{equation}\nonumber
    \evaprate = -2\frac{v_r \Sigma}{R}>0,
\end{equation}
where the final inequality follows from the fact that $v_r$ is negative (see Equation~\ref{eq:radial velocity}).
As a direct result of using the approximation Equation~\eqref{eq:MM_rad_approx}, we find that  $\evaprate$ is always constrained to be positive, that is the numerical solution from the MM model always has mass evaporating from the disk into the corona.

Considering on the other hand our vertically integrated model, and using the power-law scalings Equation~\eqref{eq:powerlaw}, the continuity Equation~\eqref{eq:cont_eq} becomes
\begin{equation}
    \evaprate=\left(p+q+\frac{3}{2}\right)\frac{v_r\Sigma}{R}.
\end{equation}
The magnitude and sign of the mass flux $\evaprate$ from the thin disk to the corona depend on the power-law indices $p$ and $q$ of $T(R)$ and $\Sigma(R)$. Using the solutions for these indices for the cooling-dominated ($r<r_b$) and evaporation-dominated ($r>r_b$) regimes, as discussed in Section~\ref{sec:two_regimes}, we find
\begin{equation}\label{eq:mdot_analytic}
\evaprate=
    \begin{cases}
        +\frac{3}{2}\frac{v_r\Sigma}{R}<0 & (r<r_b),\\
        -\frac{3}{2}\frac{v_r\Sigma}{R}>0&(r>r_b).
    \end{cases}
\end{equation}
We see that disk matter evaporates into the corona ($\evaprate>0$) in the outer regime ($r>r_b$). Moreover, the coefficient $-3/2$ is not very different from the coefficient $-2$ in the ansatz Equation~\eqref{eq:MM_rad_approx}. The MM model thus ought to give reasonable results in this regime. 

However, for the cooling-dominated regime ($r<r_b$), $\evaprate$ is negative, i.e., gas from the corona condenses back onto the thin disk. This regime would appear to be inconsistent with the framework of the MM model. Nevertheless, the MM model does produce a change in the radial dependencies of quantities at about the correct break radius (see Figure~3 in \mlmh), and the logarithmic slopes at radii below the break appear to have similar values to those predicted by our vertically integrated model. Presumably, the MM model is able to achieve this, even though it assumes evaporation at all radii, because at radii below the break the cooling term $Q^-$ dominates over the evaporation term $Q^{\rm evap}$ and so choosing the wrong sign for the latter term causes only a minor error.

\section{Transition between disk-corona system and hot accretion flow}\label{sec:truncation_radius}

Using our analytic solutions, we now consider the state transition problem. We begin by defining the mass accretion rate in the corona,
\begin{equation}
    \dot{M}_c\equiv 2\int \rho (-v_r)2\pi R\,{\rm dz}\approx -4\pi R \Sigma v_r,
\end{equation}
where the extra factor of $2$ is because we consider the total coronal accretion rate in both corona layers, whereas the $\Sigma$ we defined earlier corresponds to only one side. When expressed in units of the Eddington mass accretion rate (defined with a fiducial radiative efficiency of $0.1$),
\begin{equation}\nonumber
    \dot{M}_{\rm Edd}=\frac{L_{\rm Edd}}{0.1 c^2}=\frac{40\pi GM m_p}{\sigma_T c},
\end{equation}
the dimensionless coronal accretion rates in the two analytical solutions are
\begin{equation}\label{eq:analytic_corona_accrate}
\dot{m}_c(r)\equiv\frac{\dot{M}_c(r)}{\dot{M}_{\rm Edd}}=
    \begin{cases}
    0.034\, \kappacorr^{-\frac{1}{2}}\,\newalpha^3\,\pcorr^4\, \left(\frac{r}{r_b}\right)^{\frac{3}{2}} &(r<r_b),\\
    0.046 \,\kappacorr^{-\frac{1}{2}}\,\newalpha^3\,\pcorr^4\,\left(\frac{r}{r_b}\right)^{-\frac{3}{2}} & (r>r_b).
    \end{cases}
\end{equation}
As before, we see that the results are independent of the BH mass. The coefficients on the two sides differ by about 30\%, suggesting that neither of the two analytical solutions is particularly accurate near $r=r_b$. Equation~\eqref{eq:analytic_corona_accrate} shows that the radial power-law index in $\dot{m}_c$ is positive for $r<r_b$ and negative for $r>r_b$. This means that the maximum coronal accretion rate occurs at around the break radius $r_b$ (consistent with \mlmh).

Since our model ignores mass loss in a wind, the sum of the coronal mass accretion rate $\dot{M}_c(R)$ and the thin disk accretion rate $\dot{M}_d(R)$ should be independent of $R$. From Equation~\eqref{eq:analytic_corona_accrate}, the coronal accretion rate scales as $\dot{M}_c\propto R^{-3/2}$ at large $R$, which means that coronal accretion is negligible at very large distances. Defining $\dot{M}_{d,0}\equiv \dot{M}_d(\infty)$ as the disk accretion rate at asymptotically large $R$, mass conservation gives
\begin{equation}\label{eq:Mdot_sum}
    \dot{M}_c(R)+\dot{M}_d(R)=\dot{M}_{d,0} = {\rm (constant)}. 
\end{equation}
Since our model provides an estimate of $\dot{M}_c$ as a function of $R$, we can use Equation~\eqref{eq:Mdot_sum} to calculate how the accretion rate in the thin disk $\dot{M}_d$ varies with radius.

\begin{figure*}[ht]
\gridline{\fig{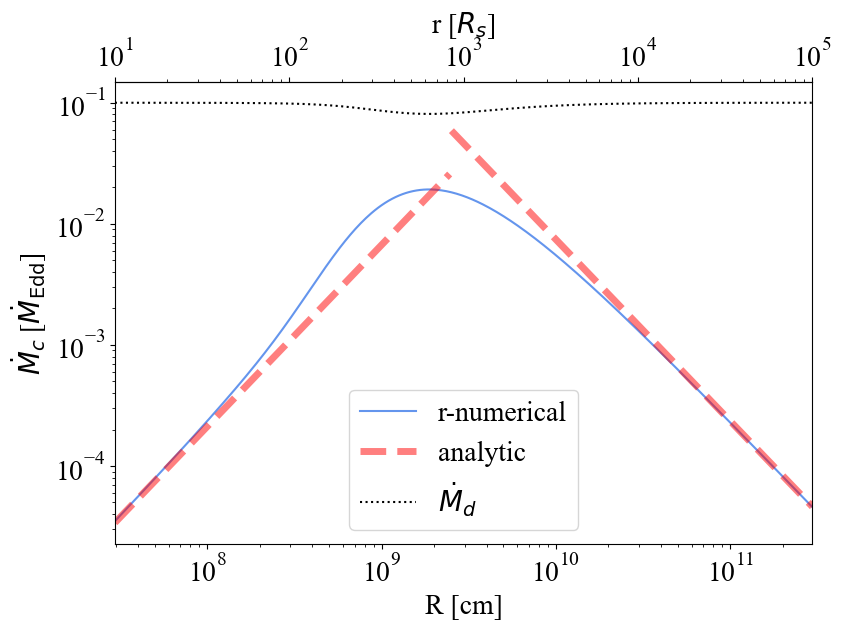}{.45\textwidth}{(a) $\dot{M}_{d,0}=10^{-1}\dot{M}_{\rm Edd}$}
\fig{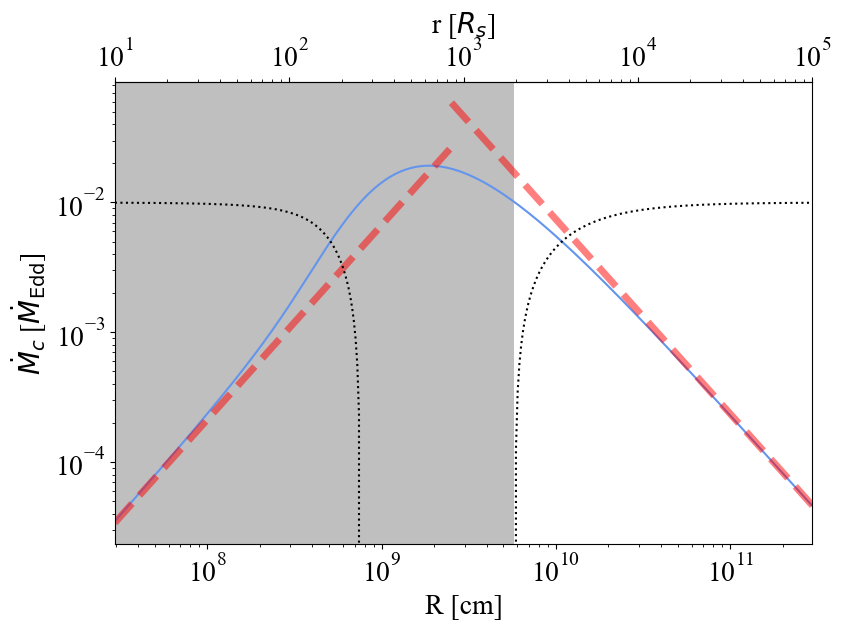}{.45\textwidth}{(b) $\dot{M}_{d,0}=10^{-2}\dot{M}_{\rm Edd}$}
}
\caption{Coronal accretion rate $\dot{M}_c(R)$ in Eddington units for two choices of the initial thin disk accretion rate, $\dot{M}_{d,0}\equiv \dot{M}_d(R \to \infty)$. The blue solid line represents numerical calculations of $\dot{M}_c(R)$, and the red dashed lines represent analytic results in the two regimes. The dotted black line shows the Eddington-scaled thin disk accretion rate $\dot{M}_d(R)$. \emph{(a)} Shows a model in which $\dot{M}_{d,0}$ is 10\% of the Eddington accretion rate. In this case, the thin disk survives at all radii, and the model corresponds to the thermal state. \emph{(b)} Model with $\dot{M}_{d,0} = 1\%$ of the Eddington rate. Here, the thin disk evaporates fully and truncates at $r\approx 2000$ (which corresponds to $R\approx 10^{10}$\,cm for a $10M_\odot$ BH). At this radius, $\dot{M}_c = \dot{M}_{d,0}$, and $\dot{M}_d\to0$. Radii below this are shaded in gray to indicate that accretion here is entirely in a hot accretion flow (there is no more thin disk). This model corresponds to the hard state.
\label{fig:truncation}}
\end{figure*}

In Figure~\ref{fig:truncation}, the thin disk accretion rate, $\dot{M}_d(R)=\dot{M}_{d,0}-\dot{M}_c(R)$, is shown together with numerical and analytic coronal accretion rate estimates $\dot{M}_c$ for two values of the total mass accretion rate $\dot{M}_{d,0}$.
In panel (a), we consider the case when $\dot{M}_{d,0}$ is 10\% of the Eddington mass accretion rate. For this relatively high rate, the thin disk accretion rate $\dot{M}_d$ decreases only slightly even at the peak of the coronal accretion rate $\dot{M}_c$. The thin disk loses a modest fraction of its mass with decreasing radius until $r=r_b$ and then it gains back the mass via condensation at smaller radii. The thin disk survives at all radii, so this model corresponds to the thermal state.

From Equation~\eqref{eq:analytic_corona_accrate} and the numerical solution shown in Figure~\ref{fig:truncation}, we see that the maximum accretion rate in the corona is found at $r\sim r_b$, and is approximately equal to
\begin{equation}\label{eq:def_m_cmax}
\dot{m}_{c,{\rm max}} \approx 0.02\, \kappacorr^{-\frac{1}{2}}\,\newalpha^3\,\pcorr^4.
\end{equation}
The model shown in panel (a) in Figure~\ref{fig:truncation} corresponds to a case where the Eddington-scaled total accretion rate $\dot{M}_{d,0}/\dot{M}_{\rm Edd}=0.1$ is larger than $\dot{m}_{c,{\rm max}}
$. Disk evaporation has only a minor effect on the system.

Panel (b) shows a different example in which the total mass accretion rate is only 1\% of the Eddington rate, $\dot{M}_{d,0}/\dot{M}_{\rm Edd}=0.01$. In this case, the thin disk accretion rate $\dot{M}_d$ decreases rapidly with decreasing $r$ and goes to 0 at some radius outside $r_b$; in the example shown, this happens at 
$r\approx 2000$. Interior to this radius (gray shaded region) there is no thin disk, and accretion occurs entirely via a hot coronal flow. 
If the transition from the thermal state to the hard state is caused by a rapid decrease in $\dot{M}_{d,0}$, one could imagine a scenario in which the inner thin disk does not have enough time to respond to the decreasing accretion rate and therefore survives temporarily. In this case, there is a possibility that some gas from the hot accretion flow might condense back on the the surviving inner disk. However, the inner disk will probably eventually disappear due to viscosity \citep{Meyer2007,Liu2007}, and once this happens, we will have the hard state.

In between the two choices of $\dot{M}_{d,0}$ considered in Figure~\ref{fig:truncation}, the system under consideration will undergo a state transition if it has a time-varying $\dot{M}_{d,0}$ (possibly caused by disk instability \citep{Lasota2001} or galaxy mergers). For the particular parameters chosen here, the transition would happen when the total mass accretion rate is around 2\% of the Eddington rate.

We note an important implication of this model. Since matter condenses back onto the thin disk for $r<r_b$, if the thin disk manages to survive down to $r=r_b$ with non-zero $\dot{M}_d$, then $\dot{M}_d$ will only increase via condensation at smaller radii and so the disk-corona structure will survive all the way down to the BH. Therefore, the truncation radius where the thin disk transitions to a hot accretion flow should always be greater than $r_b$. This prediction is in some tension with observations which suggest that the truncation radius is sometimes found at smaller radii \citep[e.g.,][]{Yuan2004}. We discuss this issue further in Section~\ref{sec:discussion_truncation}.


\section{Generalized Model with Direct Magnetic Heating} \label{sec:mag}

The vertically-integrated two-zone description of the corona we have described so far is designed to match the basic version of the MM model. The main feature of the model is that it is simple and fast, while still giving similar results as the more detailed MM model (see Figure~\ref{fig:compare_znum_ran_condense} in the Appendix). Taking advantage of this simplicity, we now generalize our model further.

It is believed that, in addition to viscous heating, coronae above thin disks are also directly heated by energy transported from the underlying thin accretion disk. The transport mechanism is likely via magnetic fields and the heating mechanism is probably through wave damping and magnetic reconnection of field loops, analogous to the situation in the solar corona \citep{Haardt1991,Haardt1993,Field1993}.
We now generalize our model such that, in addition to the viscous $Q^+_{\rm visc}$ heating term Equation~\eqref{eq:def_Qplus_visc} that we previously introduced, we include a second heating channel which we call ``magnetic heating.'' For simplicity, we assume that a fixed fraction $a$ of the viscously dissipated energy $3GM\dot{M}_d/8\pi R^3 ~[{\rm erg\,cm^{-2}s^{-1}}]$ in the thin disk \citep{Frank2002} is transferred to the corona via magnetic field lines. The total coronal heating rate is then
\begin{equation}\label{eq:allheating}
    Q^+(R)=Q^+_{\rm visc}(R)+Q^+_{\rm mag}(R), \qquad Q^+_{\rm mag}(R) \equiv a \frac{3GM\dot{M}_d}{8\pi R^3}.
\end{equation}
\citet{Liu2002b,Liu2016} investigated the value of the proportionality factor $a$ ($f$ in their notation). However, it is hard to obtain a first-principles estimate of $a$, so we treat it as a free parameter, constrained merely by the requirement $0\leq a\leq1$. Even though we call the new direct heating term magnetic heating, there is no specific field strength that the model associates with this process; indeed, direct heating from the thin disk does not even have to be associated with magnetic fields (though, in analogy with the solar corona, it probably is). 

When obtaining solutions, once again we have the choice of either a numerical approach or an analytical approach. Numerically, the method is almost the same as in Section~\ref{sec:numerical_sol}, the only difference being the addition of the new heating term $Q^+_{\rm mag}$ in the right-hand side of Equation~\eqref{eq:explicit_eq_c}. The thin disk accretion rate $\dot{M}_d$ appearing in $Q^+_{\rm mag}$ is obtained as a function of radius through Equation~\eqref{eq:Mdot_sum}: $\dot{M}_d(R)=\dot{M}_{d,0}-\dot{M}_c(R)$. For the analytic solution, just as we previously solved a problem with only a single source of heating, viscous heating $Q^+_{\rm visc}$ (Section~\ref{sec:two_regimes}), we again solve the case where there is only one form of heating, now magnetic heating $Q^+_{\rm mag}$ (it is not possible to find a simple power-law solution when both heating terms are simultaneously present).

\subsection{Analytical and Numerical Solutions when Heating is Dominated by Magnetic Heating}\label{sec:sols_mag_without_compt}

\begin{figure*}[ht]
\gridline{\fig{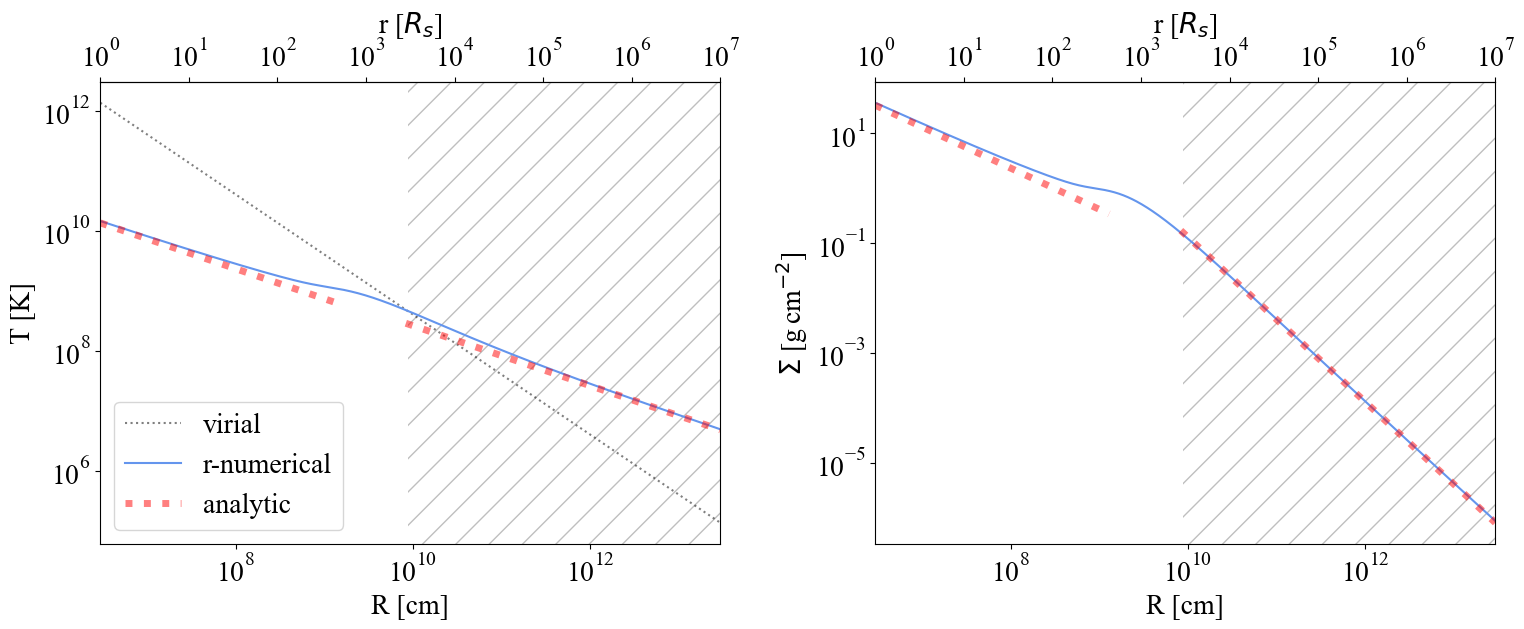}{.9\textwidth}{}
}
\caption{Numerical (blue solid line) and analytical (red dotted line) solutions when magnetic direct heating $Q^{+}_{\rm mag}$ is included. The results are for $a=0.1$ and $\dot{M}_{d,0}=0.1\dot{M}_{\rm Edd}$. In the numerical solution both $Q_{\rm mag}^+$ and $Q_{\rm visc}^+$ are included, whereas in the analytic solutions only $Q^+_{\rm mag}$ is kept. \emph{(left)} Shows the solutions for the temperature $T$. In the hatched region, the temperature exceeds the virial temperature (shown by the black dotted line). The solution here is inconsistent. \emph{(right)} Shows the surface density $\Sigma$. The relativistic temperature and high surface density in the inner radii suggests that the magnetically heated corona will be significantly affected by Compton cooling which is not considered here. The new solutions with Compton cooling are shown in Figure~\ref{fig:T_rho_magnetic}.
\label{fig:T_rho_magnetic_wo_compt}}
\end{figure*}

We begin by first identifying the radius range over which magnetic heating dominates. To do this, we use the analytical solutions obtained previously for the viscous-only heating problem (Section~\ref{sec:two_regimes}), and calculate from those solutions the ratio $Q^+_{\rm mag}(r)/Q^+_{\rm visc}(r)$.
For simplicity, we use the same fiducial parameter values as before, viz., $\newalpha= \kappacorr= \pcorr=1$, and add two new fiducial parameters, $a=0.1$, $\dot{M}_d=0.1\dot{M}_{\rm Edd}$\footnote{For simplicity, here we assume a constant thin disk accretion rate $\dot{M}_d(R)$. From Figure~\ref{fig:truncation} and Figure~\ref{fig:truncation_mag}, the thin disk accretion rate maintains its initial value $\dot{M}_{d,0}$ far from the break radius $r_b$. In the following discussion, magnetic heating dominates at radii far from $r_b$, so the assumption of constant disk accretion rate ($\dot{M}_d(R)\approx\dot{M}_{d,0}$) is valid.}. We then obtain
\begin{equation}\label{eq:ratio_mag_visc}
\frac{Q^+_{\rm mag}(r)}{Q^+_{\rm visc}(r)}\sim
    \begin{cases}
    \left(\frac{r}{10^{2.6}}\right)^{-\frac{3}{2}}&(r<r_b),\\ 
    \left(\frac{r}{10^{3.4}}\right)^{\frac{3}{2}} & (r>r_b). 
    \end{cases}
\end{equation}
We see that magnetic heating dominates over a wide range of radii, except in the vicinity of the break radius $r_b \approx 10^3$. Our previous estimate of $r_b$ (Equation~\ref{eq:r_break}) is still valid since viscous heating dominates there, and our previous analytical solutions with pure viscous heating are roughly correct in the vicinity of $r_b$ (half an order of magnitude in $r$ on either side). But everywhere else, magnetic heating dominates.

We now look for analytical power-law solutions of the model under the assumption that magnetic heating dominates. For this we substitute
\begin{equation}
    Q^+ \approx Q_{\rm mag}^+
\end{equation}
in Equation~\eqref{eq:energy_corona}. As before, the solutions have two distinct regimes. There is a regime at smaller radii where radiative cooling $Q^-$ dominates over evaporation $Q^{\rm evap}$, and where the advection term $Q^{\rm adv}$ is also negligible. Ignoring the less-important terms, we obtain an analytical solution. The second regime is at larger radii, where evaporation cooling is more important than radiative cooling, and where we obtain a different power-law solution. We do not write down the analytical solutions here because they suffer from an inconsistency which we discuss below.

The red dotted lines in Figure~\ref{fig:T_rho_magnetic_wo_compt} show the analytical solutions for the temperature $T$ and surface density $\Sigma$ for the magnetic heating-dominated problem considered in the present subsection. We plot the analytical solutions only over the range $r<10^{2.6}$ and $r>10^{3.4}$, omitting the region around $r_b\approx10^3$ where viscous heating dominates.

We see an immediate problem with the large $r$ solution shown in Figure~\ref{fig:T_rho_magnetic_wo_compt}, namely, the temperature exceeds the virial temperature $T_{\rm vir}$. Gas at supervirial temperature can easily escape a gravitationally bound system. Therefore, it seems that a magnetically heated corona with the fiducial parameters we have assumed will inevitably have a strong wind. The no-wind assumption that underlies our model thus breaks down. The model needs to be expanded to include wind physics before we can obtain trustworthy solutions at large $r$. We defer this to a future paper.

The small $r$ solution does not suffer from a wind problem. However, the temperature increases with decreasing radius and so does the surface density (previously both were constant, see Equations~\ref{eq:T_visc_sol}, \ref{eq:Sigma_visc_sol}). This means that, with decreasing radius, Compton cooling will become progressively more important and we will need to include this process in our radiative cooling term, which currently includes only \brem cooling. We discuss the importance of Compton cooling in the following subsection. In addition, the temperature even reaches $\gtrsim 10^{9.5}\,{\rm K}$, where two-temperature plasma physics may need to be considered \citep[e.g.,][]{Narayan1995}.

\subsection{Compton Cooling}\label{sec:compton_cooling}

At radii $r < r_b$, as $kT$ approaches $m_e c^2$, Compton cooling could potentially be important. To check this, we estimate the Compton $y$-parameter \citep{Rybicki1979},
\begin{equation}
    y=\frac{4kT}{m_e c^2}\left(1+\frac{4kT}{m_e c^2}\right)\tau_{\rm es} (1+\tau_{\rm es}),
\end{equation}
where $\tau_{\rm es}=\kappa_{\rm es} \Sigma$ is the electron scattering optical depth, with $\kappa_{\rm es}=\sigma_T n_e/\rho \approx 0.34\,{\rm cm^2\,g^{-1}}$
(for our assumed composition).

\begin{figure}[ht]
\gridline{\fig{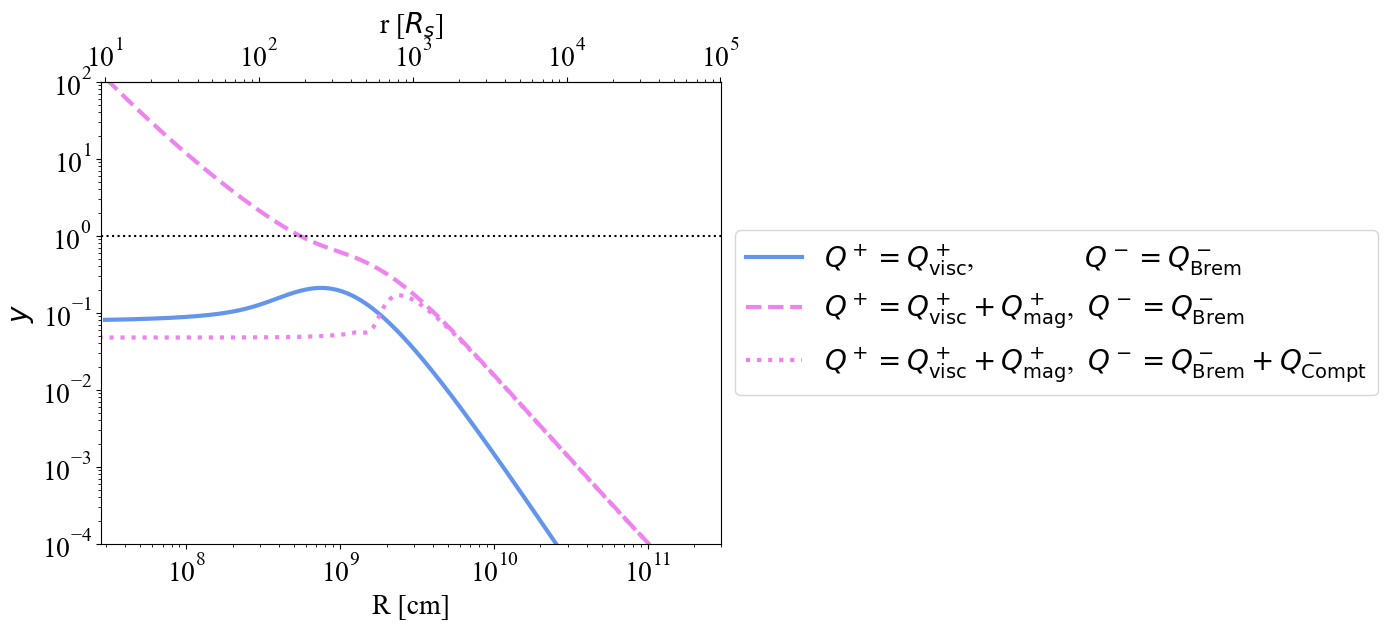}{.8\textwidth}{}
}
\caption{Compton $y$-parameter as a function of cylindrical radius $R$ as calculated from numerical solutions,  for $a=0.1$ and $\dot{M}_{d,0}=0.1\dot{M}_{\rm Edd}$. The blue solid line represents the case when heating in the corona is purely from viscosity ($Q^+=Q^+_{\rm visc}$), and the pink dashed line is when both viscous and magnetic heating are present ($Q^+=Q^+_{\rm visc}+Q^+_{\rm mag}$). The horizontal black dotted line shows $y=1$, above which Compton cooling becomes dominant. Note that when magnetic heating is included with \brem as the sole cooling mechanism, Compton cooling becomes important closer to the black hole. The new solution discussed in Sections~\ref{sec:analytic_sols_mag} and \ref{sec:energetics_mag}, where the corona is heated by both $Q^+_{\rm visc}$ and $Q^+_{\rm mag}$ and cooled by both \brem and Compton ($Q^-=Q^-_{\rm Brem}+Q^-_{\rm Compt}$), is shown by the pink dotted line. After including Compton cooling consistently, the Compton $y$-parameter now remains less than $1$ at all radii.
\label{fig:comptonY}}
\end{figure}
The Compton $y$-parameter shows a large difference between the pure viscous heating model and the magnetic heating model. In the former case (Section~\ref{sec:numerical_sol}), we find $y<1$ at all radii (see Figure~\ref{fig:comptonY}). Thus Compton cooling is less important\footnote{Note that, even when $y<1$, Compton cooling might still dominate over \brem when $\dot{m}_d$ is large. However, we did not include Compton cooling for pure viscous heating model (Section~\ref{sec:model_setup}) because we wished to match our model to early work on the MM model (e.g., \mmh, \mlmh) as closely as possible.
}.
However, with direct heating both $T$ and $\Sigma$ increase with decreasing $r$ (see Figure~\ref{fig:T_rho_magnetic_wo_compt}), and so the Compton $y$-parameter becomes very large as we approach smaller radii (Figure~\ref{fig:comptonY}). This is an indication that our neglect of Compton cooling is a serious limitation.
In reality, as the temperature tries to increase with decreasing $r$, Compton cooling (and possibly pair production) will limit the temperature to lower values than the pure \brem model predicts.

We therefore expand our model in this section to include a Compton cooling term in the equations. 
Since Equation~\eqref{eq:rad_cooling_approx} is based on the work of \citet{Johnston2017} which did not consider Compton cooling, the assumption of height-integrated \brem in the corona being equal to height-integrated \brem in the transition region should still be valid. Thus, for the transition region we use
\begin{equation}
    Q^-_t(R) \approx Q^-_{\rm Brem}(R)=n_e n f T^{1/2} H.
\end{equation}
In the corona, however, Compton cooling is important. Hence we write
\begin{equation}
    Q^-_c(R)=Q^-_{\rm Brem}(R)+Q^-_{\rm Compt}(R).
\end{equation}

Comptonization is complex and non-linear if the scattering optical depth $\tau_{\rm es}$ is large or the Compton $y$-parameter exceeds unity \citep[e.g.,][]{Rybicki1979,Sunyaev1980,Dermer1991}. In the following, we simplify the problem by assuming that both $\tau_{\rm es}$ and $y$ are reasonably small compared to unity (as we show later, this assumption is valid for our solutions). In this limit, since the soft photon energy density, $u_{\rm ph}$, of the radiation above the  photosphere of the thin disk is related to the total radiation flux, $F_{\rm d, tot}$, emerging from the disk as
\begin{equation}
    u_{\rm ph}=\frac{2}{c} F_{\rm d,tot},
\end{equation}
the Compton cooling term is 
\begin{equation}\label{eq:def_QCompt}
    Q^-_{\rm Compt}(R)\equiv \frac{4kT}{m_e c^2}\left(1+\frac{4kT}{m_e c^2}\right)  \tau_{\rm es}(1+\tau_{\rm es}) \, c \, u_{\rm ph}= 2\,y\,F_{\rm d,tot}.
\end{equation}
We write $F_{\rm d,tot}$ as the sum of the flux $F_d$ originally emitted by the thin disk and the backscattered flux $F_b$ which irradiates the disk from the corona,
\begin{equation}\label{eq:def_Fdtot}
F_{\rm d,tot}=F_d+F_b.
\end{equation}

In the case of $F_d$, we recall that a fraction $a$ of the thin disk luminosity is assumed to be transported directly to the corona. The remainder emerges as radiative flux, hence
\begin{equation}\label{eq:def_Fd}
    F_d=(1-a)\sigma T_{\rm eff, disk}^4=(1-a)\frac{3GM\dot{M}_d}{8\pi R^3}.
\end{equation}
The second equality in Equation~\eqref{eq:def_Fd} follows from standard thin disk theory: $\sigma T^4_{\rm eff, disk}=3GM\dot{M}_d/(8\pi R^3)$ \citep{Frank2002}. 

\begin{figure*}[ht]
\gridline{\fig{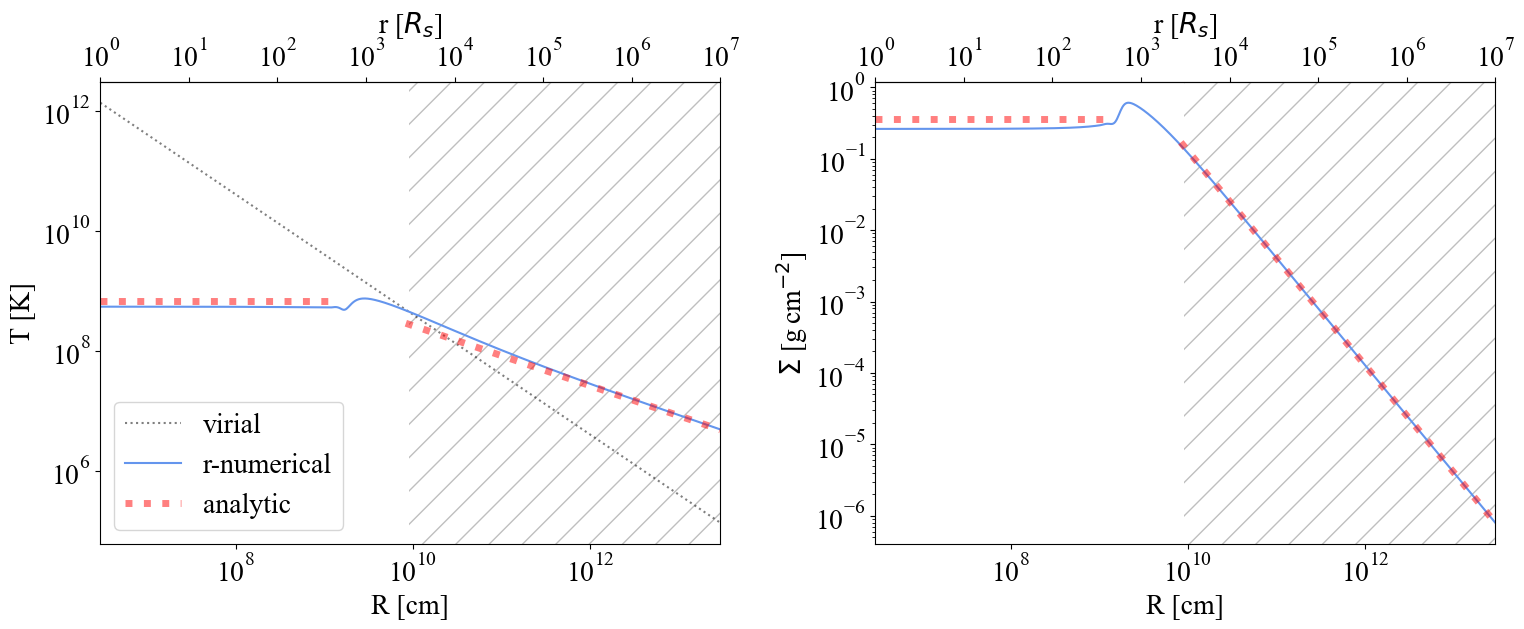}{.9\textwidth}{}
}
\caption{Similar to Figure~\ref{fig:T_rho_magnetic_wo_compt}, but with Compton cooling included. The results are again for $a=0.1$ and $\dot{M}_{d,0}=0.1\dot{M}_{\rm Edd}$. Compared to Figure~\ref{fig:T_rho_magnetic_wo_compt}, only solutions at the inner radii are modified by including Compton cooling. Both the coronal temperature $T$ \emph{(left)} and the surface density $\Sigma $ \emph{(right)} here become independent of radius. The hatched region indicates radii with supervirial temperature, where the model is inconsistent because it ignores coronal winds.
\label{fig:T_rho_magnetic}}
\end{figure*}

In the case of $F_b$, we first note that a fraction $\tau_{\rm es}(1+\tau_{\rm es})$ of the outgoing flux $F_{\rm d,tot}$ from the disk is Compton-scattered by the corona (we keep the $(1+\tau_{\rm es})$ correction factor for completeness, even though we have assumed that $\tau_{\rm es}$ is small). Each scattered photon gains in energy by a factor $[1+4kT/m_ec^2(1+4kT/m_ec^2)]$, and half of these Comptonized photons move back toward the disk surface. These photons irradiate the disk and are in part reprocessed into blackbody radiation and are in part reflected \citep{Haardt1991}. All of this irrradiation energy re-emerges from the disk with flux $F_b$. We thus write 
\begin{eqnarray} 
    F_b &=& 
    \frac{1}{2}\, \tau_{\rm es}(1+\tau_{\rm es}) \left[1 +\frac{4kT}{m_e c^2} \left( 1 +\frac{4kT}{m_e c^2} \right)\right]\,2 F_{\rm d,tot} \nonumber \\
    &=& [y+\tau_{\rm es}(1+\tau_{\rm es})] F_{\rm d, tot}.\label{eq:def_Fb}
\end{eqnarray}
We assume that the Compton-backscattered radiation remains soft and contributes fully to Compton cooling\footnote{This is a reasonable assumption because we find that the corona temperature is not relativistic in Equation~\eqref{eq:T_mag_sol} and Figure~\ref{fig:T_rho_magnetic}. The soft photon originally from the thin disk is likely to stay soft after it has Compton scattered in the sub-relativistic corona and reflected back.}. Substituting \eqref{eq:def_Fb}  in \eqref{eq:def_Fdtot}, we obtain
\begin{equation}
    F_{\rm d,tot} = \frac{F_d}{[1-y-\tau_{\rm es}(1+\tau_{\rm es})]}\,,
\end{equation}
and \eqref{eq:def_QCompt} then gives
\begin{equation}\label{eq:Compt1}
    Q^-_{\rm Compt}=\frac{2y}{[1-y-\tau_{\rm es}(1+\tau_{\rm es})]}F_d.
\end{equation}

Having thus obtained an expression for Compton-cooling in the corona, the new set of equations describing the model consists of
\begin{eqnarray}\label{eq:explicit_eq_cont_mag}
    \frac{1}{R}\dv{\,}{R}\left(R\Sigma v_r \right) &=& \evaprate ~~ {\rm (continuity)},\\
    Q^{\rm evap}_t(R) + Q^-_{\rm Brem}(R) &=& Q^{\rm cond}(R) ~~{\rm (transition~region)},\\\label{eq:explicit_eq_c_mag}
    Q^{\rm adv} (R) + Q^{\rm evap}_c(R) &=& Q^+_{\rm visc}(R)+Q^+_{\rm mag}(R) - Q^-_{\rm Brem}(R)-Q^-_{\rm Compt}(R) - Q^{\rm cond}(R) ~~{\rm (corona)}.
\end{eqnarray}

\subsection{Analytical Solutions when Heating is Dominated by Magnetic Heating and Compton Cooling is included}\label{sec:analytic_sols_mag}

When Compton cooling is included and magnetic heating dominates, the analytical solution in the evaporation-dominated regime (outer radii) remains unchanged from the solution shown in Figure~\ref{fig:T_rho_magnetic_wo_compt}. This is because cooling is negligible at these radii, so it does not matter how we model the cooling. 

However, Compton cooling strongly modifies the cooling-dominated regime (inner radii). Since Compton cooling dominates in the corona, we use the following approximate equations to calculate the analytical solution at smaller radii,
\begin{eqnarray}\label{eq:energy_balance_tr_magCompt}
    Q^{\rm cond}(R) &\approx& Q^-_{\rm Brem}(R)  ~~{\rm (transition~region, ~cooling~dominated)},\\ \label{eq:energy_balance_c_magCompt}
    Q^+_{\rm mag}(R) &\approx& Q^-_{\rm Compt}(R) ~~{\rm (corona, ~cooling~dominated)}.
\end{eqnarray}
In the transition region, the downward conductive energy flux balances \brem radiative cooling, while in the corona, direct magnetic heating balances Compton radiative cooling. All other energy terms are negligible, as confirmed by the numerical results shown in Figure~\ref{fig:energy_magnetic}.
Furthermore, since it turns out that the temperature is not relativistic ($T\sim {\rm few}\times 10^8\,{\rm K}$ (see below)) and the corona is optically thin $\tau_{\rm es}<1$,
we use the simpler expression, 
\begin{equation}\label{eq:def_y_analytic}
    y_{\rm analytic}\approx \frac{4kT}{m_e c^2} \,\tau_{\rm es},
\end{equation}
and since $y_{\rm analytic }\lesssim 0.1$, we use
\begin{equation}\label{eq:Compt2}
    Q^-_{\rm Compt, analytic}\approx 2 \, y_{\rm analytic}\, F_d, 
\end{equation}
when deriving the analytical solution.

The full analytical solutions are given in the Appendix Section~\ref{sec:appendix_magnetic}. Here we show the solutions as a function of $r/r_b$ for the temperature, 
\begin{equation}\label{eq:T_mag_sol}
T=
    \begin{cases}
    6.4\times 10^8\,\kappacorr^{-\frac{1}{5}}\,\left(\frac{\newa}{1-a}\right)^{\frac{2}{5}}\,{\rm K}&(r<r_b),\\
    4.8\times 10^8 \,\kappacorr^{-\frac{5}{6}}\,\newalpha\,\left(\newa\,\newmdotdnaught\right)^{\frac{1}{3}}\,\pcorr^{\frac{5}{3}} \left(\frac{r}{r_b}\right)^{-\frac{1}{2}}\,{\rm K} & (r>r_b),
    \end{cases}
\end{equation}
the surface density,
\begin{equation}
\Sigma =
    \begin{cases}
    0.34 \,\kappacorr^{\frac{1}{5}}\,\left(\frac{\newa}{1-a}\right)^{\frac{3}{5}}\,{\rm g\,cm^{-2}}&(r<r_b),\\
    
    0.80 \, \kappacorr^{-\frac{5}{6}}\,\newalpha^2\, \left(\newa\, \newmdotdnaught\right)^{\frac{1}{3}}\,\pcorr^{\frac{19}{6}}\left(\frac{r}{r_b}\right)^{-\frac{3}{2}}\,{\rm g\,cm^{-2}} & (r>r_b),
    \end{cases}
\end{equation}
and the average density,
\begin{equation}
\Bar{\rho} =
    \begin{cases}
    2.6\times 10^{-9}\,\kappacorr^{-\frac{6}{5}}\,\newalpha^{3}\,\left(\frac{\newa}{1-a}\right)^{\frac{2}{5}}\,m^{-1}\,\pcorr^{4}\left(\frac{r}{r_b}\right)^{-\frac{3}{2}} \,{\rm g\,cm^{-3}}&(r<r_b), \\
    7.2\times 10^{-9}\,\kappacorr^{-\frac{23}{12}}\,\newalpha^{\frac{9}{2}}\, \left(\newa\,\newmdotdnaught\right)^{\frac{1}{6}}\,m^{-1} \,\pcorr^{\frac{19}{3}} \left(\frac{r}{r_b}\right)^{-\frac{11}{4}} \,{\rm g\,cm^{-3}} &(r>r_b).
    \end{cases}
\end{equation}
In the above solutions, $\newa\equiv a/0.1$ and $\newmdotdnaught\equiv \dot{M}_{d,0}/0.1\dot{M}_{\rm Edd}$. 
Note that the solutions for both the temperature and the surface density are mass-independent, just as with the viscous heating dominated solutions (Section~\ref{sec:two_regimes}). The analytical solutions for the coronal temperature $T$ and surface density $\Sigma$ are shown as red doted lines in Figure~\ref{fig:T_rho_magnetic}. The results are consistent with observations of Seyfert 1 AGN coronae in that the mean temperature in these systems is measured to be $kT = 65\pm 10\,{\rm keV}$ or $T \sim 7.5\times 10^8\,{\rm K}$ \citep{Akylas2021} and no strong correlation has been observed between coronal temperature and black hole mass \citep{Kamraj2022}.
From the above analytical solutions, we also obtain the coronal accretion rate in Eddington units,
\begin{equation}
\dot{m}_c(r)=
    \begin{cases}
    0.035\,\kappacorr^{\frac{3}{2}}\,\newalpha^{-2}\, \frac{\newa}{1-a} \,\pcorr^{-\frac{7}{2}}\left(\frac{r}{r_b}\right)^{\frac{3}{2}} &(r<r_b),\\
        
    0.061 \,\kappacorr^{-\frac{1}{6}}\,\newalpha\,\left(\newa\,\newmdotdnaught\right)^{\frac{2}{3}}\,\pcorr^{\frac{4}{3}}\left(\frac{r}{r_b}\right)^{-\frac{1}{2}} & (r>r_b).
    \end{cases}
\end{equation}
Correspondingly, the red dotted lines in Figure~\ref{fig:truncation_mag} show the analytical solutions for $\dot{m}_c$.
Finally we note that $4kT/m_e c^2 < 1$ and $\tau_{\rm es} < 1$ at all radii, hence the Compton $y$-parameter always stays below unity. This is confirmed by the numerical result shown in Figure~\ref{fig:comptonY}.

We note that, while the numerical model discussed in the next subsection is valid for any value of $a$ over the range $0\leq a\leq 1$, the analytical solutions described here are consistent only when $a$ is relatively small. This is because, as $a$ approaches $1$, the Compton cooling $Q^-_{\rm Compt}$ is suppressed by the $(1-a)$ factor (see Equations~\ref{eq:def_Fd} and \ref{eq:Compt2}), while the direct heating $Q^+_{\rm mag}$ is enhanced by a factor $a$. The result is that the corona in the model heats up to a relativistic temperature and the approximation in Equation~\eqref{eq:def_y_analytic} breaks down. In addition, the Compton $y$-parameter approaches $1$ so the simpler form of $Q^-_{\rm Compt}$ in Equation~\eqref{eq:Compt2} is not valid (since $(1-y-\tau(1+\tau))^{-1}\not\approx1$). For accurate results, the full expression for $Q^-_{\rm Compt}$ is required, and we need the numerical solution. However, the analytical solution presented here is within a factor of $2$ of the numerical solution so long as $a\leq 0.5$, which is a reasonable range for $a$.

\subsection{Numerical Solution and Energetics}\label{sec:energetics_mag}

\begin{figure}[ht]
\gridline{\fig{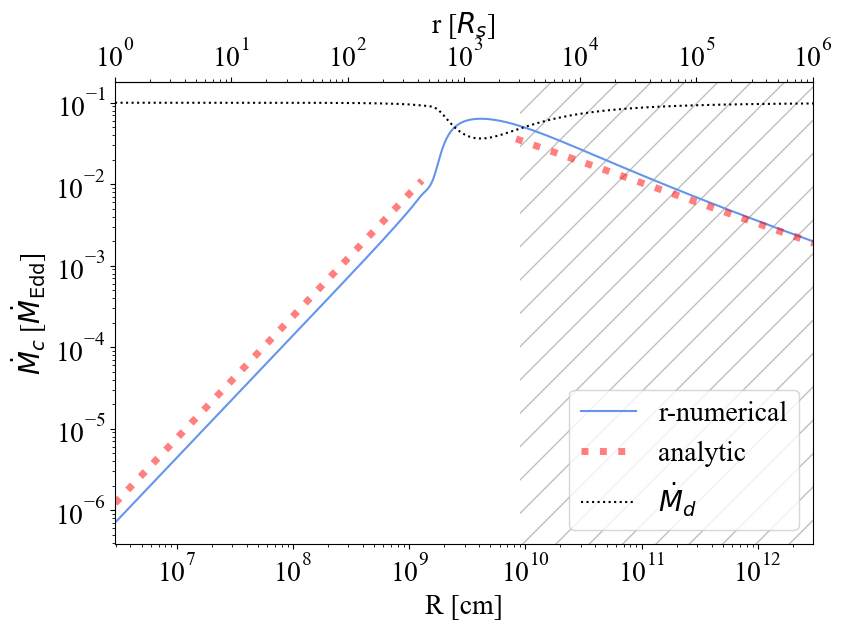}{.5\textwidth}{}
}
\caption{Similar to Figure~\ref{fig:truncation}, but for a model which includes direct magnetic heating and Compton cooling. The numerical (blue solid line) and analytical (red dotted lines) solutions for the coronal mass accretion rate $\dot{M}_c$ are shown for $a=0.1$ and $\dot{M}_{d,0}=0.1\dot{M}_{\rm Edd}$. The thin disk accretion rate $\dot{M}_d$ is shown as a black dotted line.
The hatched region shows radii at which the temperature in the solution is supervirial.
\label{fig:truncation_mag}}
\end{figure}

The blue solid lines in Figures~\ref{fig:T_rho_magnetic} and \ref{fig:truncation_mag} show the results we obtain by numerically integrating Equations~\eqref{eq:explicit_eq_cont_mag}-\eqref{eq:explicit_eq_c_mag}, supplemented with Equation~\eqref{eq:Mdot_sum}. In the corona energy equation we have included both the viscous and magnetic heating terms (Equation~\eqref{eq:allheating}), and we have set $a=0.1$ and $\dot{M}_{d,0}=0.1\dot{M}_{\rm Edd}$, as for the analytical solutions. The numerical solutions agree fairly well with the analytical solutions, the agreement becoming better the farther we are from the break radius $r_b$. At radii $r>r_b$, the numerically computed temperature is even larger than the analytical solution, so the presence of a strong wind is even more certain. The small offset between the analytical (red dotted) and numerical (blue) lines at radii below $r_b$ is because the analytical model uses the simpler version of Compton cooling in Equation~\eqref{eq:Compt2}, whereas the numerical calculation uses the full expression in Equation~\eqref{eq:Compt1}

Figure~\ref{fig:truncation_mag} indicates that a model with an initial thin disk accretion rate $\dot{M}_{d,0}=0.1\dot{M}_{\rm Edd}$ does not evaporate fully when we include magnetic heating with $a=0.1$ (just as in the previous case when only viscous heating was considered, see Figure~\ref{fig:truncation}). The maximum accretion rate in the corona for the case of Figure~\ref{fig:truncation_mag} is $\dot{m}_{c,{\rm max}}\approx 0.064$, but it is subject to change with different choices of $a$ and $\dot{M}_{d,0}$. Magnetic heating causes the critical accretion rate below which the disk truncates to increase to around $\dot{m}_{\rm crit}\sim 0.05$ for the fiducial set of parameters, $\kappacorr=\newalpha=\newa=\pcorr=1$ (compare with the critical rate $\dot{m}_{\rm crit}\sim 0.02$ for the pure viscous heating model in Equation~\ref{eq:def_m_cmax}). This revised estimate of $\dot{m}_{\rm crit}$ should however be treated with caution since the regime we are considering likely has a strong wind in the outer radii, which we have not included in our model.

\begin{figure}[ht]
\gridline{\fig{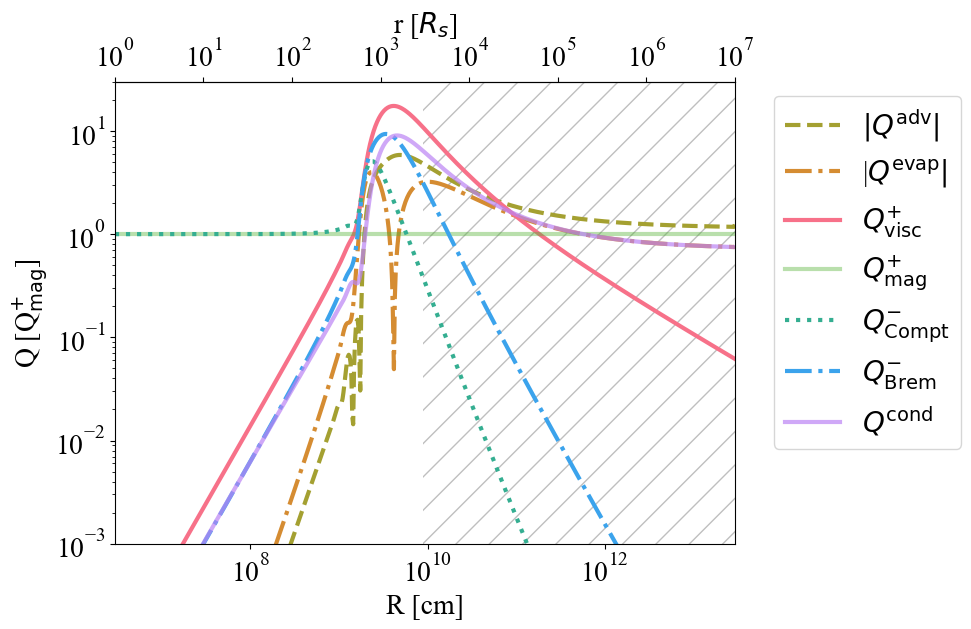}{.7\textwidth}{}
}
\caption{Energy terms normalized by the direct magnetic heating term $Q^+_{\rm mag}$ for the numerical solution (blue lines) shown in Figures~\ref{fig:T_rho_magnetic} and \ref{fig:truncation_mag}.
The hatched region indicates radii with supervirial temperature.
\label{fig:energy_magnetic}}
\end{figure}

Figure~\ref{fig:energy_magnetic} shows the variation of the different energy terms with radius, as determined from the numerical solution. The plot illustrates the complexity of the system when we include direct magnetic heating and Compton cooling.
As predicted by the analytic work, viscous heating $Q^+_{\rm visc}$ dominates for radii near $r_b \approx 10^3$, and magnetic heating dominates everywhere else. 
For $r<r_b$, the only dominant terms are magnetic heating $Q^+_{\rm mag}$ and Compton cooling $Q^-_{\rm Compt}$, which corresponds to the energy balance in the corona (Equation~\ref{eq:energy_balance_c_magCompt}). Also note that in the inner radii \brem cooling and conduction balance each other $Q^-_{\rm Brem}\approx Q^{\rm cond}$, which is the energy balance in the transition region (Equation~\ref{eq:energy_balance_tr_magCompt}). 
For $r>r_b$, evaporation $Q^{\rm evap}$ dominates over both radiative cooling terms $Q^-_{\rm Brem}$ and $Q^-_{\rm Compt}$. All of these results are consistent with the assumptions we used in deriving the analytic solutions.

\section{Discussion}\label{sec:discussion}

\subsection{Location of the Disk Truncation Radius}\label{sec:discussion_truncation}
Observational estimates of the thin disk truncation radius in hard state BHs
\citep[e.g.,][]{Yuan2004} show a wide range of radii from $10-10^4 R_S$. In the case of two low-luminosity active galactic nuclei (LLAGN), \citet{Quataert1999} argued that the truncation radius is at around $10^2 R_S$. According to our model, however, the thin disk can only truncate at a radius greater than the break radius $r_b$ (Section~\ref{sec:truncation_radius}), which 
is $\approx10^3$ for our fiducial set of parameters, $\alpha_{0.3}=\Tilde{\kappa}=\Tilde{p}=1$. As Equation~\eqref{eq:r_break} shows, $r_b$ would be smaller than $10^3$ for other values of the parameters. For example, \citet{Rosner1989} and \citet{Narayan2001} pointed out that the effective conduction coefficient in a collisionless turbulent magnetized plasma might be reduced by up to a factor of 5, i.e., $\kappacorr \approx 1/5$. This will cause $r_b$ to decrease by a similar factor. Additionally, our fiducial choice $\Tilde{p}=1$ assumes that the pressure is entirely from the thermal gas. If there is a non-negligible magnetic pressure, then as \citet{Meyer2002} noted, we will have $\pcorr=(1+1/\beta)$ (Equation~\ref{eq:pcorr_to_beta}), where $\beta$ is the plasma-$\beta$ parameter. Shearing box 3D simulations of the magnetorotational instability show that the viscosity parameter $\alpha$ and the plasma-$\beta$ are tightly correlated in differentially rotating disks such that $\alpha \beta \sim 0.5$ \citep{Blackman2008,Hawley2011}. For our fiducial choice $\alpha=0.3$, this gives $\beta\sim 1.67$ and a pressure correction factor $\pcorr=1.6$. Since $r_b$ scales as $\Tilde{p}^{-3}$, this reduces the break radius by a factor $\sim 4$. Making $\alpha$ larger can produce even more extreme results, but our fiducial value of 0.3 is already rather large, and going higher would be unreasonable.

\citet{Yuan2004,Cabanac2009} showed that for X-ray luminosities of BHBs $L_X\gtrsim 10^{-2}L_{\rm Edd}$, the truncation radius $R_{\rm tr}\lesssim 10 R_S$, and as $L_X$ declines from $10^{-2}L_{\rm Edd}$ to $10^{-3}L_{\rm Edd}$, $R_{\rm tr}$ recedes farther and farther away from the ISCO. These observations are qualitatively consistent with our model (Section~\ref{sec:truncation_radius}). The critical accretion rate above which the hard state disappears in our model when only viscous heating is present is $\dot{m}_{c, {\rm max}}\approx 0.02$ Eddington, and the limit is a factor of a few larger when we include magnetic heating (though predictions in this case are unreliable because the corona is likely to have a strong wind and we do not include this in the model). Furthermore, the analysis in Section~\ref{sec:truncation_radius} shows that the truncation radius should increase with decreasing mass accretion rate, and this is qualitatively in agreement with observations.

\subsection{Strong Coronal Outflow}\label{sec:outflow}
Strong accretion disk winds have been observed in many BHBs in the thermal state \citep{Ponti2012}.
In the best-studied example, GRO J1655-40 \citep{Miller2006,Miller2008}, the wind is launched at a radius of $\approx 10^3 R_S$. From the discussion of direct magnetic heating in Section~\ref{sec:analytic_sols_mag}, our model predicts for fiducial parameters that winds should be present for $r>r_b\approx10^3$, but not at smaller radii. Because the wind is not included self-consistently in our model, we cannot estimate the amount of mass loss in the wind. However, considering the highly supervirial temperature of our no-wind model, one suspects that the winds in these systems will be quite heavy.

Similar winds are thought to be present also in AGN. Most AGN spectra show broad emission lines originating from what is called the broad line region (BLR) in the vicinity of the SMBH. A popular model is that the BLR lines are produced by outflowing gas in a disk wind (e.g. \citealt{Lu2019,Matthews2020}, and references therein). \citet{Kollatschny2013,Kollatschny2013b,Matthews2020} estimated the wind launching radius to be $\sim 10^3 R_S$, which is consistent with our model.

If the wind launching radius is around the break radius $\sim r_b$ as our model predicts, then  Equation~\eqref{eq:r_break} indicates an explicit dependence of the launching radius on the values of the parameters, $\alpha$, $\kappacorr$, $\pcorr$, $a$, etc. In principle, it may be possible to use observations of disk winds to constrain these parameters. This might become practical if an extension of the current model that includes winds is developed.

\subsection{Comparisons with Other Models}\label{sec:other_models}

Various numerical models have been proposed based on the original work of \mmh (see \citealt{Liu2022} for a review). 
In later works, the same authors discussed the effect of changing the viscosity parameter \citep{Meyer-Hofmeister2001,Qiao2009} and reducing the conductivity of the plasma \citep{Meyer-Hofmeister2006}. A virtue of our analytic solutions is that they show explicitly how various quantities depend on the viscosity parameter and conduction coefficient. Our model is consistent with the previous numerical studies. 
An improved prescription for radial derivatives was introduced in  \citet{Meyer-Hofmeister2003} in place of the approximation shown in Equation~\eqref{eq:MM_rad_approx}. However, they still limited themselves to evaporation, whereas our model shows that there is condensation at radii $r<r_b$. 

Building upon the MM evaporation model, \citet{Rozanska2000a,Rozanska2000b,Liu2002,Qian2007} considered the effects of a two-temperature plasma and Compton cooling at small radii in the accretion system. A scenario where the Compton-cooled corona irradiates the underlying disk was studied by \citet{Liu2011,Qiao2017}, which is similar to our model set-up in Section~\ref{sec:compton_cooling}. However, our model is different in that we additionally consider direct magnetic heating.
Recently, \citet{Cheng2020} further improved the model by including radiation pressure and magnetic-reconnection heating in the model, and compared the predictions with observations of broadband spectra.  The work presented here is different from these studies in that we provide simple and analytical solutions. Quantitative comparisons will be possible only after our model is expanded to include two-temperature effects.

Our disk-corona model is complementary to the work of \citet{Spruit2002} who investigated the evaporative process and transition to ADAF specifically at the truncation radius. Re-condensation of the ADAF back to an inner disk region closer to the black hole has been studied by \citet{Liu2006,Liu2007,Meyer2007,Taam2008} by starting from ADAF solutions and modifying them to consider disk-corona interactions.

\subsection{Limitations of the Current Model}

In \S\ref{sec:assumptions}, we listed a number of approximations we have made to derive the vertically-integrated disk-corona model presented in this paper. The resulting model is sufficiently simple that it permits us to obtain analytic solutions. These solutions reveal how the properties of the corona depend explicitly on parameters, and also how the physics varies between different regimes, e.g., evaporation-dominated vs. cooling-dominated regimes. While these are significant advantages, the model does have some limitations which demand more work.

At large radii $r>r_b$, the version of our model with direct magnetic heating (Section~\ref{sec:mag}), which we view as being more realistic than the simpler pure viscous heating version (Section~\ref{sec:radial_analytic}), predicts supervirial temperatures. This is a sign that the model is not self-consistent. In practice, such systems will release the excess thermal energy by driving a wind to infinity, but our model explicitly assumes that there is no wind. We will need to modify the boundary conditions at the upper surface of the corona zone to allow mass and energy loss. If this could be accomplished without making the model overly complicated, it would allow a deeper understanding of disk winds in BHBs and AGN BLRs (see Section~\ref{sec:outflow}).

At radii $r < r_b$,
the ion and electron temperatures might deviate from each other, and two-temperature plasma physics effects may need to be considered. Energy transfer from ions to electrons could become inefficient, which might lead to a decrease in the conductive electron heat flux into the transition region and thus a decrease in the mass evaporation rate. A detailed investigation of this effect is left for future work.
Additionally, if the magnetic field is strong enough, synchrotron cooling and synchrotron self-Compton (SSC) effects could become important as well. Simple one-zone prescriptions for synchrotron cooling have been developed for the ADAF model \citep{Narayan1995}, and some of these may be of value for the corona problem.

In summary, the previous two paragraphs show that, when we include direct magnetic heating, there are serious problems at large radii (winds become important) and potential problems at small radii (two-temperature physics may need to be considered). However, the region around the break radius $r_b$ appears to be represented reasonably well by the present model, so our estimate of $r_b \approx 10^3\,\kappacorr\, \newalpha^{-2}\,\pcorr^{-3}$ (Equation~\ref{eq:r_break}) is probably safe.

Finally, we note that general relativistic gravity should replace our Newtonian gravity at sufficiently small radii, say $r \lesssim 10$.

\section{Summary}\label{sec:conclusion}

The present work was inspired by the disk evaporation model proposed by \citet{Meyer1994} and \citet{Meyer2000} to explain state transitions in black hole accretion disks. Their model, which we refer to as the MM model, considers a classical thin cold disk in the mid-plane, sandwiched between two hot coronae above and below it. Heat conduction from the the corona toward the thin disk causes gas in the thin disk to evaporate into the corona. The model calculates the structure of the corona and the mass evaporation rate by numerically solving a set of ordinary differential equations in the vertical direction. If the evaporation rate is large enough, the thin disk is truncated at a certain radius, and accretion proceeds via a hot accretion flow at smaller radii. This corresponds to the hard state (see also \citealt{Esin1997}). Alternatively, if evaporation is unable to eliminate the disk completely, the thin disk extends down to the black hole, and we obtain the thermal state.

In the present paper, we described a vertically-integrated version of the MM disk evaporation model, which allows both analytical and numerical exploration of the radial structure of the corona. Motivated by the distinct two-zone vertical structure of numerical solutions of the MM model (see Figure~\ref{fig:vertical_meyer}), we derived separate height-integrated equations for a vertically extended corona zone and a narrow transition zone between the corona and the thin disk. Assuming that cooling is dominated by \brem emission, heating is purely by viscous dissipation (both as in the MM model), neglecting mass loss in a wind (the MM model includes winds), and applying conservation laws, the model reduces to three fundamental equations. These are Equations~\eqref{eq:cont_eq}, \eqref{eq:energy_tr}, and \eqref{eq:energy_corona}, two of which are differential equations in radius, and one is algebraic. The main results of this pure viscous heating version of the model are as follows.
\begin{itemize}
\item 
For a fixed dimensionless radius $r$ scaled to the Schwarzschild radius of the black hole, the solutions for the temperature $T$~[K] of the corona, the vertically integrated surface density $\Sigma~[{\rm g\,cm^{-2}}]$, and the Eddington-scaled coronal mass accretion rate $\dot{m}_c$, are all independent of the black hole mass $M$ (the volume average density $\Bar{\rho}~[{\rm g\,cm^{-3}}]$ scales inversely with $M$). The model thus predicts that both corona formation and state transitions should be similar in stellar-mass and supermassive black hole systems, in rough agreement with observations.
\item 
With suitable approximations, we obtain analytical solutions for the radial structure of the corona. As a function of $r$, we find two well-defined regions which are separated at a break radius, $r_b\approx10^3\, (\alpha/0.3)^{-2}$, where $\alpha$ is the viscosity parameter in the corona. Equation~\eqref{eq:r_break} shows the scaling of $r_b$ with other model parameters.
\item
For $r>r_b$, the analytical solution gives $T\propto r^{-1}$, $\Sigma\propto r^{-2}$, $\dot{m}_c\propto r^{-3/2}$. We call this the evaporation-dominated region because the physics of the transition zone is dominated by the energetics of evaporation. For $r<r_b$ (cooling-dominated regime), we find that $T$ and $\Sigma$ are independent of $r$, and $\dot{m}_c\propto r^{3/2}$. In this region, gas condenses back from the corona onto the thin disk, and radiative cooling dominates the energetics of the transition zone. These analytical solutions agree with the results of detailed vertical structure solutions computed along the lines of the MM model (Figure~\ref{fig:compare_znum_ran_condense}). The full analytical solutions are presented in Sections~\ref{sec:two_regimes}, \ref{sec:truncation_radius}, and Appendix~\ref{sec:appendix_viscous}.
\item 
Exact solutions as a function of radius, with no approximations, can be easily calculated numerically. They agree well with the analytical solutions in the respective regions (Figure~\ref{fig:T_rho_viscous}). 
\item 
The maximum mass accretion rate in the corona is $\dot{m}_{c,{\rm max}} \approx 0.02\, (\alpha/0.3)^{3}$ (Figure~\ref{fig:truncation}). If a thin accretion disk has an accretion rate $\dot{M}_{d,0}/\dot{M}_{\rm Edd} < \dot{m}_{c,{\rm max}}$ at large $r \gg r_b$, the disk will be truncated and the system will be in the hard state. If $\dot{M}_{d,0}/\dot{M}_{\rm Edd} > \dot{m}_{c,{\rm max}}$, the thin disk will extend down to the black hole and the system will be in the thermal state.
\item
The model predicts that in the hard state the disk truncation radius is always far from the black hole at $r>r_b$. This disagrees with apparently smaller truncation radii observed in some systems (Section~\ref{sec:discussion_truncation}).
\item
The solutions of this version of the vertically-integrated model are consistent with our use of pure \brem cooling and our neglect of mass loss in a wind. The model is thus fairly self-consistent at all radii.
\end{itemize}

Apart from providing analytical solutions, physical understanding, and an easy route for calculating numerical solutions, the vertically-integrated model has another advantage over the original MM model. It is simple to include additional physical effects in the model. In Section~\ref{sec:mag}, we included a second source of heating (on top of viscous heating), namely, direct transfer of heat from the thin disk to the corona (presumably via magnetic fields). We find that with this additional heating, Compton cooling becomes important. We therefore include Compton cooling in addition to \brem cooling. The resulting equations are nearly as simple as those of the pure viscous heating model and give the following results.
\begin{itemize}
\item
The solutions for $T$, $\Sigma$ and $\dot{m}_c$ continue to be independent of the black hole mass, and the break radius remains roughly the same as before, $r_b\approx10^3\, (\alpha/0.3)^{-2}$.
\item 
For $r>r_b$, in regions where direct heating dominates over viscous heating, we find an analytical solution in which $T\propto r^{-1/2}$, $\Sigma\propto r^{-3/2}$, $\dot{m}_c\propto r^{-1/2}$. The full analytical solutions are presented in Section~\ref{sec:analytic_sols_mag} and Appendix~\ref{sec:appendix_magnetic}. In this solution, $T$ exceeds the virial temperature, which suggests that there should be strong mass loss in a wind (Section~\ref{sec:outflow}). The heaviest mass loss is likely to be at radii $r \sim r_b$, in agreement with observations of disk winds in stellar-mass black holes and the broad-line region in supermassive black holes. However, the model here is inconsistent since it does not include the effect of winds.
\item
For $r<r_b$, we find another analytical solution (Section~\ref{sec:analytic_sols_mag}, Appendix~\ref{sec:appendix_magnetic}) with $T\sim 5\times 10^8\,{\rm K}$, $\Sigma\sim 0.3\,{\rm g\,cm^{-2}}$, $\dot{m}_c\propto r^{3/2}$. Here, the cooling of the coronal gas is almost entirely via Compton scattering.
\item
The critical accretion rate which separates the hard state (at lower $\dot{m}$) and the thermal state (at higher $\dot{m}$), increases to $\dot{m}_{\rm crit}\approx 0.05$ for $\alpha=0.3$, $\kappacorr=\pcorr=1$, $a=0.1$, but this result should be taken with caution.
\end{itemize}

The inconsistency identified in the second bullet point of the direct-heating model indicates that further improvements are needed; specifically, we need to include coronal winds at large radii. Given the simplicity of the framework, we anticipate that such an improvement will be tractable.

\acknowledgments
We thank Martin Elvis, Jonathan Grindlay, John Raymond and Aneta Siemiginowska for useful discussions and suggestions. We also thank the referee for their careful review and for suggesting improvements to the manuscript. This work was supported in part by NSF grants PIRE OISE-1743747 and AST-1816420. The authors acknowledge support from the Gordon and Betty Moore Foundation and the John Templeton Foundation via grants to the Black Hole Initiative at Harvard University.

\software{Astropy \citep{2013A&A...558A..33A},
        NumPy \citep{vanderWalt2011NumPy},
        Matplotlib \citep{Hunter2007Matplotlib}
        }

\appendix

\section{The MM Disk Evaporation Model}\label{sec:appendix_meyer}

\subsection{Equations}\label{sec:appendix_meyer_eq}
The numerical description of disk corona evaporation model first proposed by \mmh involves solving differential equations in the vertical direction at a given radius. The full solution as a function of radius is obtained by repeating the vertical structure calculation at multiple radii and stitching the solutions together. In 3D cylindrical coordinates, they assume axisymmetry and introduce an approximation for radial derivatives to reduce the equations to a set of ordinary differential equations, only leaving $z$-derivatives. Specifically, the approximation for the radial derivatives of some quantity $X$ is
\begin{equation}\nonumber
    \frac{1}{R}\frac{\partial}{\partial R} (RX)\to -\frac{2}{R}X,
\end{equation}
where one example for the mass conservation equation was shown in Equation~\eqref{eq:MM_rad_approx}.
The mass conservation Equation~\eqref{eq:mass_cons} is then modified to
\begin{eqnarray}\label{eq:meyer_cont_eq}
    \dv[]{}{z}\,(\rho v_z)=\frac{2}{R}\rho v_r-\frac{2z}{R^2+z^2}\rho v_z.
\end{eqnarray}
The very last term on the right-hand side is added to self-consistently model the spherically expanding wind. Also, the $z$-component of the momentum equation simplifies to
\begin{equation}
    \rho v_z \dv{v_z}{z}=-\dv{P}{z}-\rho\frac{GMz}{(R^2+z^2)^{3/2}}.
\end{equation}
As described in Section~\ref{sec:eq_momentum}, the radial and azimuthal momentum equations are used to define the angular velocity $\Omega$ and radial velocity $v_r$ respectively. The energy equation is modified with the similar approximation of the radial derivative to become
\begin{equation}\label{eq:meyer_energy_eq}
    \dv{}{z}\,\left[v_z\epsilon + F_c\right]=\frac{3}{2}\alpha P \Omega -n_e n \Lambda(T)+\frac{2}{R}v_r \epsilon-\frac{2z}{R^2+z^2}[v_z\epsilon+F_c],
\end{equation}
where 
\begin{equation}\nonumber
    \epsilon=\rho\frac{v^2}{2}+\frac{\gamma}{\gamma-1}P-\rho \frac{GM}{(R^2+z^2)^{1/2}}
\end{equation}
is the energy density. Instead of \brem, they used radiative cooling function $\Lambda(T)$ of an optically thin low-density plasma from \citet{Raymond1976}. The last term in the Equation~\eqref{eq:meyer_energy_eq} is again the due to the spherically expanding wind. Finally, the last equation is the conduction equation
\begin{equation}\label{eq:meyer_cond_eq}
    F_c=-\kappa_0 T^{5/2}\dv{T}{z}.
\end{equation}
One can numerically solve for four unknowns $P$, $T$, $v_z$, $F_c$ as functions of $z$ with four differential Equations~\eqref{eq:meyer_cont_eq}-\eqref{eq:meyer_cond_eq} and the following boundary conditions.

\subsection{Boundary Conditions}\label{sec:appendix_meyer_bc}

In the four boundary conditions below, the subscript `$0$' indicates quantities at the bottom of the transition region ($z=z_d$ in our notation) and the subscript `$1$' indicates quantities at the top of the corona $z=H$.
\begin{itemize}
    \item Sonic transition at the top of the corona \\$v_z(z=H)=v_{z,1}=c_s$
    \item No influx of heat at the top of the corona \\ $F_c(z=H)=F_{c,1}=0$
    \item Chromospheric temperature at the bottom of the transition region \\ $T(z=z_d)=T_0=10^{6.5}\,{\rm K}\ll T(z=H)=T_1$
    \item Small heat inflow at the bottom of the transition region proportional to the pressure \\ $|F_c(z=z_d)=F_{c,0}|\propto P_0\ll \rm{max}(|F_{c}|)$
\end{itemize}
It is a free boundary problem where the height of the corona ($H$) is set to be at the sonic point ($v_z=c_s$).

\subsection{Vertical Numerical Solutions for No-wind Case}\label{sec:appendix_vertical_no_wind}

In Section~\ref{sec:nowind_approx}, we compared our no-wind approximated model with the MM model.
Here we describe the method used to obtain this no-wind solution, which is slightly modified from the MM model to be consistent with our vertically-integrated two zone model. We also confirm in this subsection that the obtained no-wind numerical solutions are in good agreement with our two-zone analytic solutions in Section~\ref{sec:two_regimes}.

In detail, the wind terms (the last term in Equations~\ref{eq:meyer_cont_eq} and \ref{eq:meyer_energy_eq}) are neglected and the first boundary condition in Section~\ref{sec:appendix_meyer_bc} is adjusted such that the vertical speed is zero at scale-height $v_z(z=H)=0$. 
As we observed from Section~\ref{sec:interaction_disk_corona}, the radial derivative approximation of the MM model is not valid for $r<r_b$. Therefore for the vertical numerical solutions here, we use new simplifications that are consistent with Equation~\eqref{eq:mdot_analytic}. Then the continuity equation becomes (with the wind term neglected)
\begin{equation}\label{eq:cont_eq_discontinous}
\dv{(\rho v_z)}{z}=-\frac{1}{R}\frac{\partial}{\partial R}(R\rho v_r)=
    \begin{cases}
        -\frac{3}{2}\frac{v_r\rho}{R} & (r<r_b),\\
        +\frac{3}{2}\frac{v_r\rho}{R} &(r>r_b),
    \end{cases}
\end{equation}
and the energy equation is also similarly tuned according to analytic scaling relations.
The scale-height $H$ is set when the pressure is half of the pressure at the bottom of the transition region $P(z=H)=P_0/2$.
The lower boundary to start the numerical calculation is $z_d= R/300$.

\begin{figure}[ht]
\gridline{\fig{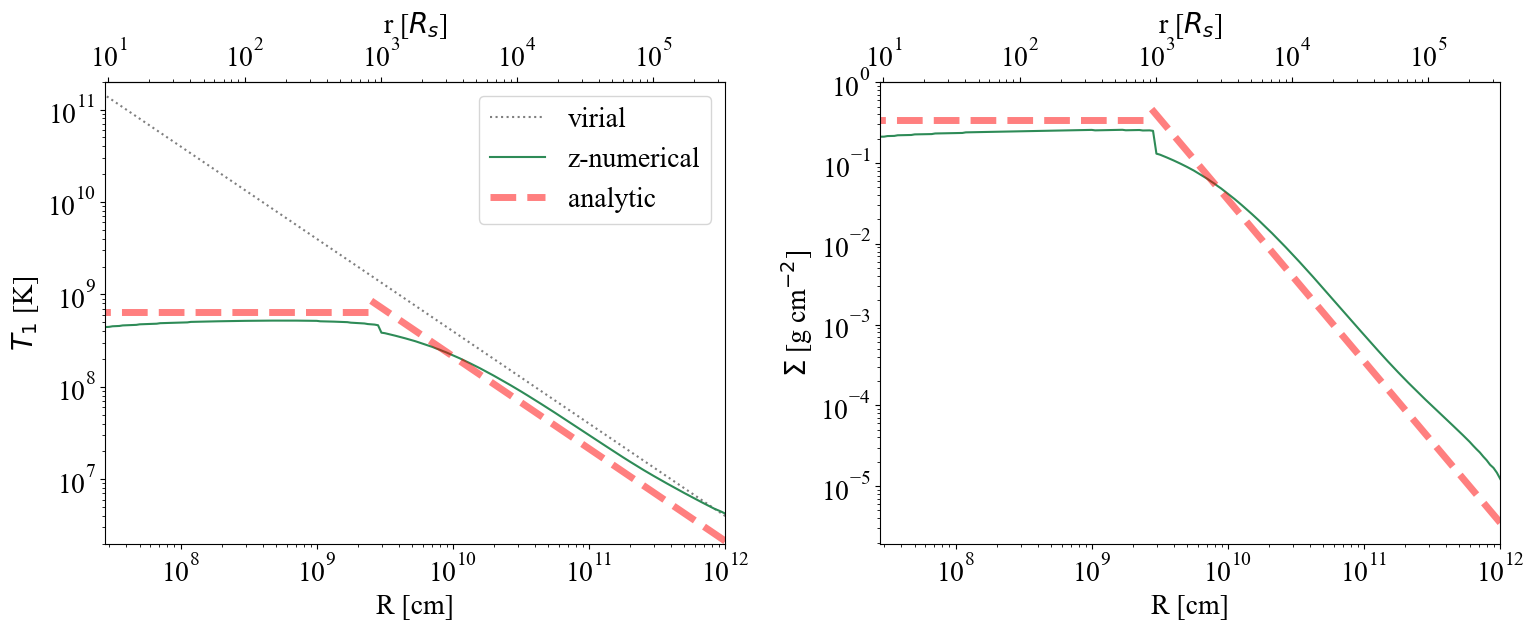}{.9\textwidth}{}
}
\caption{Comparison between analytic solutions (Section~\ref{sec:two_regimes}, red dashed lines) and no-wind vertical numerical solutions (Section~\ref{sec:appendix_vertical_no_wind}, green solid line) for \emph{(left)} coronal temperature $T_1$ and \emph{(right)} surface density $\Sigma$.
\label{fig:compare_znum_ran_condense}}
\end{figure}

The obtained no-wind numerical solutions (green solid line) are shown in Figure~\ref{fig:compare_znum_ran_condense} along with the two-zone analytic solutions (red dashed lines) for the temperature and the average density. From the vertical numerical solutions, the surface density is approximately $\Sigma\approx \rho_1 \sqrt{kT_1/\mu}/\Omega_K$ at each $R$ so this quantity is plotted in panel (b). Because of the discontinuity at $r_b$ in the Equation~\eqref{eq:cont_eq_discontinous}, there is a noticeable discontinuity around $r_b$ in numerical solutions. Note that similar equations are used for obtaining both solutions, but the applied simplifications are different. The no-wind numerical solutions are solved by making an approximation on radial derivatives (simplifying in $R-$direction similarly with the MM model) and the analytic solutions are obtained by vertically integrating the equations (simplifying in $z-$direction). The two solutions are in a very good agreement, meaning that our simple height-integration approach is reasonable and can reproduce the detailed calculation of the MM model.

Note that at large $R$, the numerical solutions diverge from the analytic solutions. This is because as $R$ increases, coronal temperature decreases and approaches the chromospheric temperature of $T_0=10^{6.5}\,{\rm K}$ set at the bottom of the transition region (Section~\ref{sec:appendix_meyer_bc}) as a boundary condition.


\section{Exact forms of analytic solutions}
The exact forms of analytic solutions are presented in this section. The solutions look slightly different here because in the main text, the dimensionless radius $r$ is normalized by the dimensionless break radius $r_b$. Each analytic solutions is presented in both regimes, with the first one for the inner cooling-dominated regime ($r<r_b$) and the second one for the outer evaporation-dominated regime ($r>r_b$).

\subsection{Viscous heating only}\label{sec:appendix_viscous}
The analytic solutions correspond to Section~\ref{sec:two_regimes} with viscosity as the only source of heating $Q^+=Q^+_{\rm visc}$.
\begin{equation}
T\approx
    \begin{cases}
    3.01\frac{m_p^2 k^3}{\kappa f \mu^3}\alpha^2\pcorr^{3}&\propto \kappa^{-1}\alpha^2\pcorr^{3} \\
    0.75\frac{\gamma-1}{\gamma+3}\frac{\mu c^2}{kr}&\propto r^{-1}
    \end{cases}
\end{equation}

\begin{equation}
\Sigma=
    \begin{cases}
        3.4\frac{m_p^4 \alpha^3}{f^2\kappa}\left(\frac{k}{\mu}\pcorr\right)^{\frac{9}{2}} &\propto \kappa^{-1}\alpha^3\pcorr^{\frac{9}{2}} \\
        
        7.1\times 10^{-2}\, \frac{c^4 (\gamma-1)^2\kappa}{\gamma\alpha(\gamma+3)}\left(\frac{\mu}{k}\right)^\frac{7}{2}\pcorr^{-\frac{3}{2}}r^{-2}&\propto \kappa\alpha^{-1} \pcorr^{-\frac{3}{2}}r^{-2}
    
    \end{cases}
\end{equation}

\begin{equation}
\Bar{\rho}=
    \begin{cases}
        0.70\frac{m_p^3 c^3 \alpha^2}{GM}\sqrt{\frac{k^5}{\mu^5 f^3\kappa}}\pcorr^{\frac{5}{2}}r^{-\frac{3}{2}} &\propto \kappa^{-\frac{1}{2}}\alpha^2 M^{-1}\pcorr^{\frac{5}{2}}r^{-\frac{3}{2}} \\
        
    2.9\times 10^{-2}\frac{(\gamma-1)^{3/2}}{\gamma (\gamma+3)^{1/2}}\frac{\kappa}{\alpha}\left(\frac{\mu}{k}\right)^{\frac{7}{2}}\frac{c^6}{GMr^3}\pcorr^{-2}&\propto \kappa\alpha^{-1} M^{-1}\pcorr^{-2}r^{-3}  
    
    \end{cases}
\end{equation}

\begin{equation}
r_b=
    \begin{cases}
    0.083\,\frac{\gamma-1}{\gamma}\frac{f\kappa c^2}{m_p^2\alpha^2}\left(\frac{\mu}{k}\right)^4\pcorr^{-3}&\propto \kappa \alpha^{-2}\pcorr^{-3} \\
        
    0.028\,\frac{(\gamma-1)(\gamma+3)}{\gamma^2}\frac{f\kappa c^2}{m_p^2 \alpha^2}\left(\frac{\mu}{k}\right)^4\pcorr^{-3}&\propto \kappa \alpha^{-2}\pcorr^{-3}
    \end{cases}
\end{equation}

%
%

\begin{equation}
\frac{\dot{M}_c}{\dot{M}_{\rm Edd}}=
    \begin{cases}
    2.9\, m_p^5\sigma_T\frac{\alpha^6}{f^3\kappa^2 c^2}\left(\frac{k}{\mu}\pcorr\right)^{\frac{17}{2}}r^{\frac{3}{2}}&\propto \kappa^{-2}\alpha^6\pcorr^{\frac{17}{2}} r^{\frac{3}{2}}\\
        
    1.5\times 10^{-2} \,\frac{\sigma_T}{m_p}\frac{(\gamma-1)^3}{\gamma(\gamma+3)^2}\left(\frac{\mu}{k}\right)^{\frac{7}{2}}\kappa c^4\pcorr^{-\frac{1}{2}}r^{-\frac{3}{2}}&\propto \kappa \pcorr^{-\frac{1}{2}}r^{-\frac{3}{2}} 
    \end{cases}
\end{equation}

\subsection{Magnetic direct heating dominated}\label{sec:appendix_magnetic}
The analytic solutions correspond to Section~\ref{sec:analytic_sols_mag} when magnetic heating dominates $Q^+\approx Q^+_{\rm mag}$ and when Compton cooling is included.
\begin{equation}
T=
    \begin{cases}
    0.55\,\left(\frac{a^2 c^4 f m_e^2}{(1-a)^2 k^2\sigma_T^2\kappa}\right)^{\frac{1}{5}}
    &\propto \kappa^{-\frac{1}{5}}a^{\frac{2}{5}}\\
    
    1.2\,\left(\frac{\gamma}{4-\gamma}\right)^{\frac{1}{3}}\left(\frac{k}{\mu}\pcorr\right)^{\frac{1}{6}}\left(\frac{m_p c^2}{\kappa\sigma_T}\right)^{\frac{1}{3}}\left(a \newmdotdnaught\right)^{\frac{1}{3}} r^{-\frac{1}{2}} 
    &\propto \kappa^{-\frac{1}{3}}a^{\frac{1}{3}}\newmdotdnaught^{\frac{1}{3}} \pcorr^{\frac{1}{6}} r^{-\frac{1}{2}}
    \end{cases}
\end{equation}

\begin{equation}
\Sigma =
    \begin{cases}
    0.27\,m_p\left(\frac{a^3 c^{6} m_e^3 \kappa}{(1-a)^3 f k^3 \sigma_T^3}\right)^{\frac{1}{5}}
    &\propto \kappa^{\frac{1}{5}} a^{\frac{3}{5}}\\
    
    0.35\,\frac{\gamma-1}{\gamma}\left(\frac{\gamma}{4-\gamma}\right)^{\frac{1}{3}}\left(\frac{\mu}{k}\right)^{\frac{7}{3}}\frac{\kappa^{2/3}}{\alpha}\left(\frac{m_p}{\sigma_T}\right)^{\frac{1}{3}}c^{\frac{8}{3}}\left(a\newmdotdnaught\right)^{\frac{1}{3}}\pcorr^{-\frac{4}{3}}r^{-\frac{3}{2}}
    &\propto \alpha^{-1} \kappa^{\frac{2}{3}} a^{\frac{1}{3}} \newmdotdnaught^{\frac{1}{3}}\pcorr^{-\frac{4}{3}}r^{-\frac{3}{2}}
    \end{cases}
\end{equation}

\begin{equation}
\Bar{\rho} =
    \begin{cases}
    0.13\,m_p (GM)^{-1}\left(\frac{a^2 c^{19} m_e^2}{(1-a)^2\sigma_T^2}\right)^{\frac{1}{5}}\left(\frac{\kappa^3}{f^3 k^{9}}\right)^{\frac{1}{10}}\left(\frac{\mu}{\pcorr}\right)^{\frac{1}{2}}r^{-\frac{3}{2}}
    &\propto M^{-1}\kappa^{\frac{3}{10}}a^{\frac{2}{5}}\pcorr^{-\frac{1}{2}}r^{-\frac{3}{2}}\\
    
    0.11\,\frac{\gamma-1}{\gamma}\left(\frac{\gamma}{4-\gamma}\right)^{\frac{1}{6}}\left(\frac{\mu}{k}\right)^{\frac{35}{12}}(GM)^{-1}\frac{\kappa^{5/6}}{\alpha}\left(\frac{m_p}{\sigma_T}\right)^{\frac{1}{6}}c^{\frac{16}{3}}\left(a\newmdotdnaught\right)^{\frac{1}{6}}\pcorr^{-\frac{23}{12}} r^{-\frac{11}{4}}
    &\propto \alpha^{-1} M^{-1} \kappa^{\frac{5}{6}} a^{\frac{1}{6}} \newmdotdnaught^{\frac{1}{6}}\pcorr^{-\frac{23}{12}}r^{-\frac{11}{4}}
    \end{cases}
\end{equation}



\begin{equation}
\frac{\dot{M}_c}{\dot{M}_{\rm Edd}}=
    \begin{cases}
    0.042\,\frac{\pcorr a m_e \alpha}{(1-a) \mu} r^{\frac{3}{2}}
    &\propto \alpha a \pcorr r^{\frac{3}{2}}\\
        
    0.12\,\frac{\gamma-1}{\gamma}\left[\frac{\gamma}{4-\gamma}\right]^{\frac{2}{3}}\left[\frac{\sigma_T c^4}{m_p}\right]^{\frac{1}{3}}\kappa^{\frac{1}{3}}\left(\frac{\mu}{k}\right)^{\frac{7}{6}}\left(a\newmdotdnaught\right)^{\frac{2}{3}}\pcorr^{-\frac{1}{6}}r^{-\frac{1}{2}}
    &\propto \kappa^{\frac{1}{3}} a^{\frac{2}{3}}\newmdotdnaught^{\frac{2}{3}}\pcorr^{-\frac{1}{6}}r^{-\frac{1}{2}}
    \end{cases}
\end{equation}

\section{cooling function}\label{sec:appendix_cooling}

In this section we discuss validity of using a \brem cooling function in our analytic approach (Section~\ref{sec:model_setup}) instead of a more realistic cooling function. In numerical studies of the MM model, the cooling function of a optically thin low-density plasma \citep{Raymond1976} was used. In Figure~1 of \citet{Raymond1976}, the cooling function $\Lambda(T)$ follows a \brem curve above $T\gtrsim 5\times 10^7\,{\rm K}$.

As shown in Figure~\ref{fig:T_rho_viscous}(a) or Figure~\ref{fig:T_rho_magnetic}(a), the coronal temperature interior to $r_b$ always have high temperature ($T\gtrsim 10^8\,{\rm K}$) regardless of which heating mechanism dominates. Thus, our assumption is reasonable in the inner regime ($r<r_b$) since cooling function asymptotes to \brem curve at such high temperatures. However, exterior to $r_b$, the temperature can get as low as $\lesssim 10^{7}\,{\rm K}$ very far from the black hole (for example, $r\gtrsim 10^5$ in Figure~\ref{fig:T_rho_viscous}(a)). For the lower temperatures, atomic processes start to contribute to the cooling function so in principle realistic cooling curve should be used instead of a simple \brem. However, analyses of energy (Figure~\ref{fig:energy_viscous} and Figure~\ref{fig:energy_magnetic}) showed that for $r>r_b$, the radiative cooling ($Q^{-}_{\rm Brem}$ or $Q^-_{\rm Compt}$) becomes negligible compared to other energy terms so the choice of cooling functions is not as important. Therefore, we conclude it is reasonable to use \brem as the only radiative cooling mechanism, except in the close vicinity of the BH in magnetically heated model where Compton cooling becomes important.


\section{Saturated conduction}
The conductive heating flux formula in Equation~\eqref{eq:def_cond_flux_classical} is taken from the classical theory of thermal conduction.
However, the theory is based on an assumption that the mean free path $\lambda$ is much smaller than the temperature scale length $L_T$. Therefore, the classical conduction cannot be applied in certain cases where $\lambda>L_T$ and the conduction reaches an upper bound or saturates \citep{Cowie1977,Komarov2018}. \citet{Cowie1977} states that the upper bound of the conductive flux exists because the energy flux is limited to the thermal energy $nkT$ times the thermal velocity. The mean free path of electrons mediating conduction is \citep{Cowie1977}
\begin{equation}
    \lambda=t_{\rm eq} \left(\frac{3kT_e}{m_e}\right)^{1/2},
\end{equation}
where the electron-electron equipartition time $t_{\rm eq}$ is \citep{Spitzer1962}
\begin{equation}
    t_{\rm eq}=\frac{3 m_e^{1/2}(kT_e)^{3/2}}{4\pi^{1/2}n_e e^4 \ln{\Lambda}}.
\end{equation}
The Coulomb logarithm is \citep{Spitzer1962}
\begin{equation}
    \ln{\Lambda}=\ln{\left[\frac{3}{2e^3}\left(\frac{k^3T_e^3}{\pi n_e}\right)^{1/2}\right]}.
\end{equation}
Comparing the mean free path to the scale-height of the corona $H$, $\frac{\lambda}{H}\sim 10^{-2}$ for radial numerical solutions where the viscosity is the only source of coronal heating (Section~\ref{sec:two_regimes}) and $\frac{\lambda}{H}\sim 10^{-1}$ when magnetic direct heating is added. In principle $\lambda$ should be compared with the temperature scale length $L_T$ which is approximately the height of the transition region $z_t$. The scale length $L_T$ is around $0.1 H$, so we conclude here that the classical conduction theory is marginally applicable to our system. \citet{Rozanska1999} also confirmed that there is no need to consider saturation of conduction in the context of AGN or galactic black holes.

\bibliography{corona}{}
\bibliographystyle{aasjournal}

\end{CJK}
\end{document}